\documentclass[12pt,a4paper]{article}
\pdfoutput=1

\usepackage{jcappub}
\allowdisplaybreaks

\usepackage[applemac]{inputenc}

\usepackage[margin=1in]{geometry}
\usepackage{graphicx}
\usepackage{amsfonts}
\usepackage{subfigure}
\usepackage{float}
\usepackage{hyperref}
\usepackage[dvipsnames]{xcolor}
\usepackage{amsmath}
\usepackage[normalem]{ulem}
\usepackage{empheq}
\usepackage{amssymb}
\usepackage{tikz}
\usepackage{tabularx}
\newcommand{\class}{{\sc class}}

\usepackage{booktabs}

\def\lsim{~\rlap{$<$}{\lower 1.0ex\hbox{$\sim$}}}
\def\bsim{~\rlap{$>$}{\lower 1.0ex\hbox{$\sim$}}}

\def\ln{{\rm ln}}

\def\be{\begin{equation}}
\def\ee{\end{equation}}
\def\bea{\begin{eqnarray}}
\def\eea{\end{eqnarray}}
\def\ba{\begin{align}}
\def\ea{\end{align}}
\def\bi{\begin{itemize}}
\def\ei{\end{itemize}}

\newcommand{\bn}{{\mathbf n}}

\def\mathbi#1{\textbf{\em #1}}

\def\bx{{\bf{x}}}
\def\bk{{\bf{k}}}
\def\bq{{\bf{q}}}

\def\bv{{\bf{v}}}

\def\ba{{\vec{g}}}

\newcommand{\HH}{\mathcal{H} }

\def\grad{\mathbi{$\nabla$}}

\newcommand{\ndv}{{v_{||}}}
\newcommand{\ndvtwo}{{v_{||}^{(2)}}}
\newcommand{\ndvthree}{{v_{||}^{(3)}}}

\title{The relativistic dipole and gravitational redshift on LSS}

\author[a,b]{Enea~Di~Dio,}
\author[a,b]{Uro\v s~Seljak}

\affiliation[a]{Physics Division, Lawrence Berkeley National Laboratory, Cyclotron Rd, Berkeley, CA 94720}
\affiliation[b]{Berkeley Center for Cosmological Physics and Department of Physics, University of California, Berkeley, CA 94720}

\abstract{
We compute the dipole of the galaxy correlation function at 1-loop in perturbation theory by including all the relevant relativistic contributions.
This provides a description and understanding of what the dipole truly measures, in particular in relation to the gravitational redshift effect in Large Scale Structure. 
In order to develop this perturbative approach we have computed for the first time the relevant relativistic corrections to third order in perturbation theory, including the corresponding non-linear galaxy bias model. 
This perturbative approach agrees on a wide range of scales with good accuracy with previous numerical results based on geodesic light tracing in N-body simulations.
Previous claims of gravitational redshift detection  may have neglected several relativistic effects which are comparable with the amplitude of gravitational redshift 
around 10 Mpc/h scales and which complicate the gravitational redshift interpretation of the measurement.
}

\begin{document}
\maketitle

\section{Introduction}
The two-point correlation function is a fundamental quantity to describe the statistical proprieties of Large Scale Structure (LSS) of our universe. In predicting the amplitude and the shape of the two-point correlation function we need to carefully consider that we observe galaxy clustering through photons which have travelled in a clumpy universe and which are affected by the motion and the gravitational potential of the source. In first approximation the mapping between real and redshift space is described by the so-called Kaiser Redshift-Space Distortions (RSD)~\cite{Kaiser:1987qv}. The anisotropy induced by RSD gives rise to non-vanishing quadrupole and hexadecapole (even multipoles). Beyond the standard Kaiser approximation several other effects appear in linear theory, by describing galaxy clustering in a relativistic framework~\cite{Yoo:2009,Yoo:2010,Bonvin:2011bg,Challinor:2011bk}.  
All the effects beyond the Kaiser approximation are suppressed at least by a factor $\HH/k$, where $\HH$ is the comoving Hubble parameter, or integral along the line-of-sight, such as lensing magnification. Interestingly, it has been first noticed in Ref.~\cite{McDonald:2009ud} that by considering two different galaxy populations the terms proportional to $\HH/k$ lead to an imaginary part of the cross power spectrum, or odd multipoles of the correlation function~\cite{Bonvin:2013ogt}. Indeed, peculiar velocities and gravitational redshift break the symmetry along the line of sight sourcing non-vanishing odd multipoles. Therefore, measurements of odd multipoles will provide a measurement\footnote{Due to wide-angle effects, part of the quadrupole, sourced by RSD, leaks in the dipole~\cite{Gaztanaga:2015jrs,Giusarma:2017xmh,Lepori:2017twd,Castorina:2017inr}. The amplitude of this effect is strongly sensitive to the definition of the dipole angle~\cite{Gaztanaga:2015jrs}.} of relativistic effects on the past light-cone. 

Within linear perturbation theory the dipole is sourced only by the so-called Doppler effect, while the gravitational redshift effect vanishes due to the Euler equation which governs the galaxy motion through the Equivalence Principle. In this view the measurement of the dipole of the correlation function may provide a test of the Equivalence Principle on cosmological scales, see Ref.~\cite{Bonvin:2018ckp}. In recent years different groups have claimed a detection of gravitational redshift effects on cluster scales~\cite{Wojtak:2011ia} and in LSS on even larger scales~\cite{Alam:2017izi}. These measurements have inspired
numerous discussions on what one is actually measuring~\cite{Zhao:2012st,Kaiser:2013ipa,Cai:2016ors}, and have shown that several terms of the same order of the gravitational redshift have been ignored in the cluster work. Recently this issue has been also studied through light-tracing in newtonian N-body simulation~\cite{Breton:2018wzk}, but a theoretical description and understanding in a relativistic framework is still missing.
Measurements of relativistic effects through odd multipoles of the 2-point function are limited by shot-noise and systematic effects. However, upcoming surveys, like DESI~\cite{Aghamousa:2016zmz}, Euclid~\cite{Laureijs:2011gra}, LSST~\cite{Abell:2009aa} and others, will be able to detect them with a sufficient statistical significance for a detection. From this point of view it is worth investigating a theoretical description of what the dipole of the correlation function truly measures beyond the linear regime.

The gravitational redshift is one of the first prediction of General Relativity (GR)~\cite{Einstein_1916}, and its first measurement has been performed in the late 50s~\cite{Pound_1959}. The first pioneering theoretical work in a cosmological framework date back to Ref.~\cite{1995A&A...301....6C}. The gravitational redshift provides a direct and independent measurement of the gravitational potential (and its spatial derivatives). In order to use gravitational redshift measurement to test the theory of gravity or the Equivalence principle on cosmological scales, we need to further understand what are the main sources of the dipole of the correlation function at the corresponding scales.
It is worth reminding that a measurement of the gravitational redshift probes the same metric potential which determines the motion of galaxies~\cite{Kaiser:2013ipa}. Hence, it provides a test of the Equivalence Principle. Gravitational redshift measurements can yield to a test of gravity if combined with weak lensing, which probes the Weyl potential.
While we know that the gravitational potential is exactly canceled by the acceleration of galaxy motion on linear scales~\cite{Bonvin:2011bg,Challinor:2011bk} (through Euler equation) and that the amplitude of gravitational potential on cluster scales is comparable to peculiar velocity and light cones effects~\cite{Zhao:2012st,Kaiser:2013ipa,Cai:2016ors}, we lack comprehensive description on intermediate scales. 

In our work we provide a theoretical description of the dipole of the correlation function at 1-loop within a relativistic framework. This will include both the effects of peculiar velocities and gravitational potentials. We expect the linear approximation to fail at the scales comparable to the claimed detection of gravitational redshift on LSS (around $10$ Mpc), and  we know that on linear scales the gravitational potential does not induce any effect on the dipole of the correlation function.
To achieve this task it is not enough to use the second order relativistic effects derived in Refs.~\cite{Yoo:2014sfa,Bertacca:2014dra,DiDio:2014lka}, but we need to compute as well the relevant relativistic effects to third order in perturbation theory.
Whenever possible we compare with the simulations of Ref.~\cite{Breton:2018wzk} showing an excellent agreement at any scale.
This agreement also provides a solid confirmation of the correctness of our derivation of relativistic effects to third order in perturbation theory, that we derive in this work for the first time.

The paper is organized as follows: in Section~\ref{sec:NumberCounts} we summarize the relativistic number number counts at different orders in perturbation theory, leaving its derivation in Appendix~\ref{app:second} and~\ref{app:third}. In Section~\ref{sec:1loop} we compute the dipole at 1-loop and we show our numerical results in Section~\ref{sec:numres}. In Section~\ref{sec:discussion} we discuss previous results and in Section~\ref{sec:conclusions} we conclude. We leave the extension to further effects such magnification and evolution bias to Appendix~\ref{sec:s_bias}.


\section{Number counts}
\label{sec:NumberCounts}
We define the fluctuation of galaxy number counts $\Delta$ in terms of the observable redshift $z$ and the photon direction $\bn$ as
\be
\Delta \left( \bn , z\right) \equiv \frac{n \left( \bn , z \right) - \langle n \rangle \left( z \right) }{ \langle n \rangle \left( z \right) }
\ee
where $<..>$ denotes the average over directions $\bn$ at fixed observed redshift $z$, and $n \left( \bn , z \right) = d N /dz/d\Omega$ is the number density of sources per redshift and solide angle.
We are interested in computing the next-to-leading order dipole of the relativistic galaxy number counts. This requires to go up to third order in perturbation theory,
\bea
\langle \Delta \left( \bn_1 , z_1\right) \Delta \left( \bn_2 , z_2\right) \rangle& \simeq& 
\langle \Delta^{(1)} \left( \bn_1 , z_1\right) \Delta^{(1)} \left( \bn_2 , z_2\right) \rangle
\nonumber \\
&+&
\langle \Delta^{(2)} \left( \bn_1 , z_1\right) \Delta^{(2)} \left( \bn_2 , z_2\right) \rangle
\nonumber \\
&+&
\langle \Delta^{(1)} \left( \bn_1 , z_1\right) \Delta^{(3)} \left( \bn_2 , z_2\right) \rangle+
\langle \Delta^{(3)} \left( \bn_1 , z_1\right) \Delta^{(1)} \left( \bn_2 , z_2\right) \rangle \, .
\qquad
\eea
In our work 
we will neglect terms integrated along the line of sight like lensing magnification and Integrated Sachs-Wolf (ISW) effects, supported by results of previous work in terms of dipole of the correlation function~\cite{Bonvin:2013ogt,Lepori:2017twd,Breton:2018wzk}.

We expand the correlation function\footnote{The correlation function can be expressed directly in terms of observables coordinates
$$
\xi \left( z_1 , z_2 , \cos \theta = \bn_1 \cdot \bn_2 \right) = \langle  \Delta \left( \bn_1 , z_1\right) \Delta \left( \bn_2 , z_2\right) \rangle \, .
$$
By assuming a fiducial cosmological model we can rewrite it in terms of more convenient coordinates $\left( d, \tilde \mu , r \right)$, where they denote respectively the pair separation, the angle between the line of sight and the pair separation vector, and the mean comoving distance.
For the sake of simplicity we will often omit the explicit dependence on the mean comoving distance $r$.
} in terms of multipoles as
\be
\xi \left( {\bf d} \right) = \sum_{\ell} \xi_\ell \left( d \right) L_\ell \left(-  {\bf d} \cdot \bn \right) 
\ee
where ${\bf d}$ is the separation vector between the two sources.
It is convenient to define the multipole expansion of the power spectrum through
\be
\label{def_multipole}
P_\ell \left( k \right) =  \frac{2\ell +1 }{2} \int_{-1}^1P \left( k, \mu \right) L_\ell \left( \mu \right) d \mu 
\ee
where $L_\ell $ denotes the $\ell$-order Legendre polynomials and $\mu = - \bn \cdot \hat\bk $. 
In flat-sky approximation, the multipoles of the correlation function and of the power spectrum are simply related by\footnote{The sign related to the imaginary unity $i$ depends on the Fourier convention, on the definition of $\mathbf{d}$ and the power spectrum normalization. We adopt the following Fourier convention
\be
\tilde f \left( \bk \right) = \int d^3x f\left( \bx \right) e^{i \bk \cdot \bx}\, , \qquad \qquad  f \left( \bx \right) = \int \frac{d^3k}{\left( 2 \pi \right)^3} \tilde f\left( \bk \right) e^{-i \bk \cdot \bx}\, .
\nonumber
\ee
We will consider the distance from source $A$ to source $B$, i.e.~$\mathbf{d} = \bx_B - \bx_A $, and the power spectrum given by $\langle \Delta_A \left( \bk \right) \Delta_B \left( \bk' \right) \rangle= \left( 2 \pi \right)^3 P^{AB}\left( \bk \right) \delta_D \left( \bk + \bk' \right) $.}~\cite{Hall:2016bmm} 
\be \label{xi_ell}
\xi_\ell \left( d \right) = i^\ell \int \frac{dk }{2 \pi^2} k^2 P_\ell \left( k \right) j_\ell \left( k d \right) 
\ee
where $j_\ell$ indicates the spherical Bessel function of order $\ell$.

Being interested in the dipole, we observe that only the terms in the power spectrum $P \left( k ,\mu \right)$ which exhibit odd parity in $\mu$ can lead to a non-vanishing dipole.
We remark that this is true only on flat-sky approximation and neglecting the effect of redshift evolution, as assumed through our work. As shown in Refs.~\cite{Gaztanaga:2015jrs,Giusarma:2017xmh,Lepori:2017twd,Castorina:2017inr}, beyond flat-sky approximation part of the quadrupole leaks into the dipole. Nevertheless the quadrupole, as any even multipole, is dominated by well-known Newtonian terms and we therefore do not need to derive further relativistic corrections. In Section~\ref{sec:discussion} we compare its contamination to the relativistic dipole defined through~\cite{Bonvin:2013ogt}
\be \label{eq:WA}
\xi_1^{\rm wide} \left( d \right)   = -\frac{2}{5} f \Delta b_1 \frac{d}{r} \int \frac{dk }{2 \pi^2} k^2 P \left( k  \right) j_2 \left( k d \right) 
\, ,
\ee
where $f$ denotes the growth rate and $\Delta b_1$ is the linear bias difference between two galaxy populations.
In our work we consider as relativistic corrections all the terms which are suppressed by a spatial derivative in real space, or a coefficient $\HH/k$ in Fourier space, with respect to the standard Newtonian perturbation theory. It is convenient to introduce a counting scheme in terms of the power of the coefficient $\HH/k$. While our derivation does not assume any specific theory of gravity (our results apply to any metric theory of gravity under the assumption of conservation of the number of photons) we consider the density fluctuation $\delta$ and the peculiar velocity $v $ to be parametrically related to the metric potential $\phi$ through
\be
\delta \sim \left( k/\HH \right)^2 \phi \qquad \text{and} \qquad v\sim \left( k/\HH \right) \phi
\ee
as predicted by Poisson and Euler equations. Under this scheme we can expand the number counts in terms of standard Newtonian terms $\Delta_N$ and relativistic corrections $\Delta_R$ as
\be
\Delta^{(n)} = \Delta_N^{(n)} + \Delta_R^{(n)} + \mathcal{O} \left( \left( \HH/k \right)^2 \right), \qquad \text{where} \quad \Delta_R^{(n)} \sim \left( \HH/ k \right)  \Delta_N^{(n)} \, .
\ee
Since pure Newtonian terms do not lead to a non-vanishing dipole we need to compute correlation between $\Delta_N$ and $\Delta_R$. Further relativistic corrections, that we neglect, are suppressed by an additional factor $\left( \HH/ k \right)^2$. Given that our counting scheme is based on the number of spatial derivatives, terms of the order $\left( \HH/ k \right)^2$ are expected to carry an even parity with respect to $\mu$.

In next section we summarize the results of relativistic corrections up to third order in perturbation theory.  We leave the derivation to the Appendixes~\ref{app:second} and~\ref{app:third}.

\subsection{First order}
Linear relativistic corrections to galaxy number counts have been first computed in Refs.~\cite{Yoo:2009,Yoo:2010,Bonvin:2011bg,Challinor:2011bk} and then generalised to non-flat geometries~\cite{DiDio:2016ykq} and vector perturbations~\cite{Durrer:2016jzq}. This expression reads as
\bea \label{counts1}
\Delta(\bn, z) &=& b \delta  + \HH^{-1} \partial_r \ndv 
 + (5s - 2) \int_0^{r} \frac{r - r'}{2 r r'} \Delta_{\Omega} (\Phi + \Psi) dr' \nonumber  \\
&&+ \left(\frac{\dot \HH}{\HH^2} + \frac{2-5 s}{r\HH} + 5 s - f_{\text{evo}}
\right)\ndv  \nonumber \\
&&+\left( f_{\text{evo}}  - 3\right)\HH V +(5s - 2) \Phi + \Psi + \HH^{-1} \dot \Phi +\frac{2-5s}{r} \int^{r}_0 dr' (\Phi + \Psi) \nonumber \\
&& +\left(\frac{\dot \HH}{\HH^2} + \frac{2-5 s }{r\mathcal{H}}+5s  - f_{\text{evo}}\right)\left(\Psi + \int^{r}_0 dr' \left(\dot\Phi + \dot\Psi \right)\right) \, ,   
\eea
where $D$ denotes the density fluctuation in comoving gauge, $\ndv = \bn \cdot \bv$ and $\bv$ is the peculiar velocity in Newtonian gauge, $\Phi$ and $\Psi$ are the two Bardeen potentials and $V$ is the velocity potential defined through $\bv = - {\bf\nabla}V $. A dot refers to the partial time derivative and the integrals are along the past light-cone. We define the comoving Hubble parameter with $\HH= \dot a/a$ and the comoving distance with $r$. The dark matter perturbations are related to galaxies through three bias parameters: a galaxy bias $b$, a magnification bias
\be
s= - \frac{2}{5} \left.\frac{\partial \ln \ \bar n \left( z , \ln L \right)}{\partial \ln L}\right|_{\bar L}
\ee
where $\bar L$ denotes the threshold luminosity of a given survey and $\bar n$ the background number
density,
and an evolution bias
\be
f_{\rm evo}= 3 - \left( 1 +z \right) \frac{d \ln \ \bar n}{dz}
\ee
which parametrizes the deviation from number conservation of sources in a comoving volume.

According to our counting scheme we define 
\bea
\Delta_N^{(1)} &=& b \delta + \HH^{-1} \partial_r \ndv \, , \\
\Delta_R^{(1)} &=& \left(\frac{\dot \HH}{\HH^2} + \frac{2-5 s}{r\HH} + 5 s - f_{\text{evo}}
\right)\ndv \, .
\eea
The Newtonian part simply includes density perturbations with the redshift-space distortion effect under the Kaiser approximation~\cite{Kaiser:1987qv}. The relativistic term describes the so-called Doppler effect which has been already studied in the recent literature~\cite{McDonald:2009ud,Yoo:2012,Bonvin:2013ogt,Irsic:2015nla,Gaztanaga:2015jrs,Bonvin:2015kuc,Hall:2016bmm,Bonvin:2016dze,Lepori:2017twd,Giusarma:2017xmh,Breton:2018wzk}.
We remark that in Eq.~\eqref{counts1} we have assumed that galaxies follow geodesic through the Equivalence Principle.
As a consequence, if galaxies are subject to acceleration, their light-cone velocity differs from the fixed time velocity, and at the leading order this happens to be exactly the same as the gradient of the gravitational potential, since the latter is the source of the acceleration.
As pointed out by Ref.~\cite{Bonvin:2018ckp}, relativistic effects provide a solid test of the Equivalence Principle and geodesic motion of galaxies at cosmological scales.

We remark again that we ignore the terms integrated along the line of sight. This includes also the lensing effects which is parametrically of the same order as the Newtonian terms. Nevertheless, the line of sight integral reduces its amplitude and therefore it is comparable to other terms only for surveys with a very poor redshift resolution and at very high redshift. In particular for photometric surveys it has been shown that a lack of lensing parametrisation can bias the cosmological parameter estimation~\cite{Cardona:2016qxn}, in particular for models beyond General Relativity (GR)~\cite{,Lorenz:2017iez,Villa:2017yfg}. 
In general for large radial correlations, lensing magnification becomes the largest term in the 2-point function. Hence, a correlation measurement of well-separated sources provides a measurement of such effect~\cite{Scranton:2005ci}.
Nevertheless its impact on spectroscopic surveys is negligible, as shown in terms of dipole of the correlation function in Refs.~\cite{Bonvin:2013ogt,Lepori:2017twd}.

\subsection{Second order}
Recently, several groups have derived independently the galaxy number counts to second order, including all the relativistic effects~\cite{Yoo:2014sfa,DiDio:2014lka,Bertacca:2014dra}. Since the full expression is extremely long we present here just the relevant terms which dominate the dipole. Given that there is no general consensus on the results of these differing derivations, we have re-derived the dominant relativistic corrections through a simple approach. Interested readers can look at Appendix~\ref{app:second} for the derivation.
Here we summarize the results:\footnote{We omit the magnification bias induced by the luminosity dependence of galaxy bias, for more details see Appendix~\ref{sec:s_bias}.}
\bea
\label{Delta2N}
\Delta^{(2)}_N &=&\delta_g^{(2)} + \HH^{-1} \partial_r \ndv^{(2)} 
+ \HH^{-1} \partial_r \left( \ndv  \delta_g \right)  + \HH^{-2}  \partial_r \left( \ndv \partial_r \ndv \right) \, ,
\\
\label{Delta2R}
\Delta^{(2)}_R &=&\left( \frac{\dot \HH}{\HH^2} + \frac{2 - 5s}{\HH r}+ 5s - f_{\rm evo}\right) \ndv^{(2)}
+ \left( 1 + 3 \frac{\dot \HH}{\HH^2} + \frac{4 - 5s}{\HH r} + 5s - 2 f_{\rm evo} \right) \HH^{-1} \ndv \partial_r \ndv
\nonumber \\
&&
 + \left( \frac{\dot \HH}{\HH^2} + \frac{2 - 5s}{\HH r} + 5s - f_{\rm evo} \right) \ndv \delta_g  
- \HH^{-1} \ndv \dot \delta_g + 2 \HH^{-1} v^a \partial_a \ndv
\nonumber \\
&&
 +  \HH^{-2} \psi \partial_r^2 \ndv + \HH^{-1} \psi \partial_r \delta_g - \HH^{-2} \ndv \partial_r^2 \psi \, .
\eea
The Newtonian part agrees with standard perturbation theory, as shown in Ref.~\cite{DiDio:2015bua}.
We also need to introduce a consistent galaxy bias expansion to second order, see Ref.~\cite{Desjacques:2017msa} and references therein,
\bea
\delta^{(1)}_g &=& b_1 \delta  \, , \\
\delta_g^{(2)} &=& b_1 \delta^{(2)} + \frac{1}{2} b_2 \left( \delta^2  { -  \langle \delta^2 \rangle} \right) + b_{K^2} \left( \left(K_{ij}\right)^2  { -\langle \left(K_{ij}\right)^2 \rangle }  \right) 
\nonumber\\
&=&
 b_1 \delta^{(2)} + \frac{1}{2} b_2 \delta^2   + b_{K^2}  \left(K_{ij}\right)^2  - { \sigma^2 \left( \frac{b_2}{2} + \frac{2}{3} b_{K^2} \right)}\, ,
 \eea
where\footnote{Following the short notation of Ref.~\cite{Desjacques:2017msa} we denote
$$
K^2 \equiv \left( K_{ij} \right)^2 \equiv {\rm tr} \left( K K \right) = K_{ij} K_{ji} \, .
$$}
\be
K_{ij} \left( \bk \right) = \left[ \frac{k_i k_j}{k^2} -\frac{1}{3} \delta_{ij} \right] \delta \left( \bk \right) \, 
\ee
and
\be
\sigma^2 \equiv \int \frac{d^3q}{\left( 2 \pi \right)^3} P\left( q \right) \, .
\ee
The last line of the relativistic part, Eq.~\eqref{Delta2R}, depends explicitly on the gravitational potential. Indeed, even when applying Euler equation, the observable number counts beyond the linear 
order depend on the metric perturbations at the source position.

\subsection{Third order}
We have computed the number counts to third order in perturbation theory considering all the relativistic effects suppressed by a single factor $\HH/k$ with respect to the Newtonian contribution.
The detailed calculation can be found in Appendix~\ref{app:third}, here we summarize the results we will use in the rest of the paper:
\bea
\label{Delta3N}
\Delta_N^{(3)} &=& { \delta_g^{(3)} +\frac{{\partial_r \ndvthree}}{\HH}}
{ 
+\left[ \HH^{-1} \partial_r\left(  \ndv\delta_g  \right)\right]^{(3)}
+ \left[ \HH^{-2}  \partial_r \left( \ndv \partial_r \ndv \right) \right]^{(3)}
}
\nonumber \\
&&
 + { \frac{1}{6} \HH^{-3}\partial_r^3 \ndv^3
 +\frac{1}{2} \HH^{-2} \partial_r^2 \left( \delta_g \ndv^2 \right)
 } \, ,
 \\
 \Delta_R^{(3)} &=&
 \left\{
{\left( \ndvthree + \left[\ndv \delta_g\right]^{(3)}  \right)\left(\frac{{\dot\HH}}{{\HH}^2}+\frac{2- 5s}{{\HH} r} +5s - f_{\rm evo} \right)} 
\right.
\nonumber \\
&&
+ \left( 1 + 3 \frac{\dot \HH}{\HH^2} + \frac{4- 5s}{\HH r} +5s - 2 f_{\rm evo} \right) \HH^{-1} \left[ \ndv \partial_r \ndv \right]^{(3)}
 - \HH^{-1} \left[ \ndv \dot \delta_g \right]^{(3)}
\nonumber\\
&&
\left.
 + 2 \HH^{-1} \left[ v_a \partial^a \ndv \right]^{(3)}
  +  \HH^{-2} \left[ \psi \partial_r^2 \ndv \right]^{(3)}
 + \HH^{-1} \left[ \psi \partial_r \delta_g \right]^{(3)} - \HH^{-2} \left[ \ndv \partial_r^2 \psi \right]^{(3)}
 \right\}
 \nonumber \\
 &&
 +
 {
 3 \HH^{-3} \partial_r \ndv \left( \partial_r^2 \ndv \psi - \ndv \partial_r^2 \psi \right)
 + \HH^{-3} \ndv \left(  \partial_r^3 \ndv \psi - \ndv \partial_r^3 \psi \right)
 }
   \nonumber \\
 &&
  + \HH^{-2} \psi \partial_r^2 \left( \ndv \delta_g \right) 
 - \HH^{-2} \partial_r^2\psi  \left( \ndv \delta_g \right) 
 +
 \frac{1}{3\HH^2} \partial^2_r \left( \ndv^3 \right) \left( 1 + \frac{3 \dot \HH}{\HH^2} +\frac{3}{\HH r} - \frac{3}{2} f_{\rm evo} \right)
      \nonumber \\
 &&
 +\frac{1}{2 \HH} \partial_r \left( \ndv^2 \delta \right)  \left( 1 + 3 \frac{\dot \HH}{\HH^2} +\frac{4}{\HH r}  - 2f_{\rm evo}\right)
 - \frac{1}{\HH^2} \partial_r \left( \ndv^2 \dot \delta \right)
    \nonumber \\
 &&
 - \frac{1}{2 \HH^2}\partial_r^2 \left( \ndv v^a v_a \right) - \frac{1}{\HH}\partial_r \left( \delta  v^a v_a \right) 
 +\frac{1}{2\HH} v^2\partial_r \delta
 + \frac{2}{\HH^2} \partial_r \left( \partial_a \ndv v^a \ndv \right)  \, .
 \label{Delta3R}
\eea
We remark that the terms in the curly brackets in Eq.~\eqref{Delta3R} correspond to the higher order contribution of the same operators defined in Eq.~\eqref{Delta2R}.
We also consistently extend the bias expansion to third order
\be
\delta_g^{(3)} = b_1 \delta^{(3)} + b_2 \delta \delta^{(2)} + \frac{1}{6} b_3 \delta^3 + 2  b_{K^2} K_{ij} K^{(2)}_{ij} +b_{K^3} \left( K_{ij} \right)^3 + b_{\delta K^2} \delta  \left(K_{ij}\right)^2 + b_{\rm td} O^{(3)}_{\rm td} 
\ee
where 
\be
O^{(3)}_{\rm td}  = \frac{8}{21} K_{ij} \mathcal{D}^{ij} \left( \delta^2 - \frac{3}{2} \left( K_{ij} \right)^2 \right)
\ee
and $\mathcal{D}^{ij}$ is defined through $K_{ij} = \mathcal{D}_{ij} \delta$.
Assuming that the comoving number of sources is conserved we can derive the evolution of $b_1$ and $b_2$, see Ref.~\cite{Desjacques:2016bnm}
\bea
\dot b_1 &=& \left( 1 - b_1 \right) f \HH
\\
\dot b_2 &=&\left( -2 b_2 - \frac{8}{21} + \frac{8}{21} b_1 \right) f \HH
\eea
and as well relate all other biases to $b_1$ and $b_2$
\bea
b_{K^2} &=& - \frac{2}{7} \left( b_1 -1 \right), \\
b_{\delta K^2}&=& \frac{1}{21} (7 b_1-6 b_2-7), \\
b_{K^3} &=& \frac{22 (b_1-1)}{63}, \\
b_{\rm td} &=& \frac{23}{42} \left( b_1 - 1 \right) \, .
\eea
In this way at any order $n$ in perturbation theory we have $n$ independent bias parameters.

\subsection{Re-normalization}

Combining the bias expansion and perturbation theory to third order we encounter some unphysical divergences. These can be cured by re-nomarlizing the operators and by absorbing the divergences in the physical bias parameters.

We consider the re-normalization of the following matter perturbations~\cite{Assassi:2014fva,Desjacques:2016bnm}
\bea
\left[\delta^2\right] &=& \delta^2 -  \sigma^2 \left( 1+ \frac{68}{21} \delta \right), \\
\left[\delta^3\right] &=& \delta^3 - 3 \sigma^2  \delta ,\\
\left[ \left( K_{ij}\right)^2 \right] &=& \left( K_{ij}\right)^2 -\frac{2}{3} \sigma^2 \left( 1 + \frac{68}{21} \delta \right), \\
\left[ O^{(3)}_{td} \right] &=& O^{(3)}_{td} -\frac{32}{63} \sigma^2 \delta \, .
\eea
Therefore for the galaxy density perturbations we find 
\bea
\label{ren_delta3}
\left[{\delta_g}\right] 
&=&
{\delta_g} -  \delta \sigma^2 \left( \frac{68}{21} \left(  \frac{b_2}{2} + \frac{2}{3}b_{K^2} \right) +\frac{b_3}{2} + \frac{32}{63} b_{\rm td} +\frac{2}{3} b_{\delta K^2} \right) 
\nonumber \\
&=&
{\delta_g} -  \delta \sigma^2 \left( \frac{b_3}{2}  -\frac{22 b_1}{189}+\frac{10 b_2}{7}+\frac{22}{189}\right)  \, ,
\eea
where we have defined\footnote{Form this point on we will neglect magnification and evolution biases. The results can easily be generalised to include these effects.} 
\be
\mathcal{R} = \frac{\dot\HH}{\HH^2} + \frac{2}{\HH r} \, . 
\ee
We can absorb these divergences as counterterms
 in
\bea
\label{Delta1_norm}
\Delta^{(1)}&\simeq& { \delta_g^{(1)} + \HH^{-1} \partial_r \ndv }{ +  \mathcal{R} \ndv^{(1)} } + \text{counter terms}
\nonumber \\
&=&  \left( b_1 - \sigma^2 \left(  \frac{b_3}{2}  -\frac{22 b_1}{189}+\frac{10 b_2}{7}+\frac{22}{189}\right) \right) \delta + \HH^{-1}  \partial_r  \ndv 
+ \mathcal{R} \ndv \, .
\eea

\section{1-loop corrections}
\label{sec:1loop}
Once derived the number counts to third order in perturbation theory, including the relevant relativistic effects, we can compute the next-to-leading order power spectrum. 
We write the observed number count fluctuations as
\be
\delta_s \left( \bk , t\right)  = \sum_{n =1}^\infty D_1^n \left( t \right) 
\int d^3\bk_1 ...  d^3\bk_n \left[ \delta_D \right]_n
 \left(Z_n \left( \bk_1, ..., \bk_n \right) + S_n \left( \bk_1, ..., \bk_n \right) \right) \delta_1 \left( \bk_1 \right) ...\delta_n \left( \bk_n \right)
\ee
where the kernels $Z_n$ denote the standard Newtonian part ($\Delta_N^{(n)}$), $S_n$ the leading relativistic effects ($\Delta_R^{(n)}$), $D_1$ is the linear growth factor and $\left[ \delta_D \right]_n= \delta_D \left(\bk- \bk_1 -... - \bk_n \right)$. 

The leading order power spectrum (which induces a non-vanishing dipole) is then 
\bea
\label{P11dip}
&& \hspace{-1.5cm} 
P^{(11)} \left( k ,\mu \right) =  \left[ Z_1^A \left( \bk \right) S_1^B \left( - \bk \right) +  
S_1^A \left( \bk \right) Z_1^B \left( - \bk \right)
\right] P\left( k,z \right) 
\nonumber \\
&=&
\left[-  Z_1^A \left( \bk \right) S_1^B \left(  \bk \right) +  
S_1^A \left( \bk \right) Z_1^B \left(  \bk \right)
\right] P\left( k,z \right)  = 
\nonumber
\\
&=&  
\left[ - i \Delta b_1 \mathcal{R} f \mu \frac{\HH}{k}
- 
\sigma^2 \mu \frac{\HH}{k} i  f \mathcal{R} \left({ \frac{22}{189}\Delta b_1 -\frac{10 \Delta b_2 }{7}-\frac{\Delta b_3}{2} }\right) 
\right]
P\left( k,z \right)   \, .
\eea
We remark again that the relativistic effects lead to an imaginary cross power spectrum, as pointed out by Ref.~\cite{McDonald:2009ud}.
The subscripts $A$ and $B$ denote two different galaxy populations and we have introduced the following notation
\be
\Delta b_i = b_i^A - b_i^B \qquad \text{and} \qquad \Delta b_{12} = b_2^A b_1^B - b_1^A b_2^B \,.
\ee
Clearly, being the dipole proportional to the difference of the bias coefficients, a single tracer does not lead to a non-vanishing dipole. As we will see in Eqs.~(\ref{P22dip}-\ref{P13dip}) this holds also beyond the leading order. Indeed, to have a non-vanishing dipole we need to have an asymmetry along the line of sight, which can not be sourced by a single tracer\footnote{We are neglecting evolution effects, which can effectively produce an asymmetry along the line of sight~\cite{Bonvin:2013ogt,Irsic:2015nla,Lepori:2017twd}. A non-vanishing dipole can also be sourced by geometrical wide-angle effects for a single tracer, if it is defined with respect to an angle which effectively breaks the symmetry along the radial direction~\cite{Gaztanaga:2015jrs,Castorina:2017inr}.}.
As expected, we see that the relativistic power spectrum is suppressed by factor a $\HH/k$ with respect to the matter power spectrum.
Beyond leading order we have two different contributions in the form of $P^{22} \left( k , \mu  \right)$ and $P^{13} \left( k , \mu  \right)$ and both of them are quadratic in the matter power spectrum. These can be expressed as
\bea
P^{22} \left( k , \mu  \right) &=& 2  \int \frac{d^3q}{\left( 2 \pi \right)^3} \left[ -Z_2^A \left( \bq, \bk - \bq \right) S_2^B \left(  \bq,  \bk- \bq \right) + S_2^A \left( \bq, \bk - \bq \right) Z_2^B \left( \bq,  \bk-\bq \right) \right] 
\nonumber \\
&& \qquad
P \left( q \right) P \left( | \bk - \bq | \right) \, ,
\\
P^{13} \left( k , \mu  \right) &=& 3 P(k) \int \frac{d^3q}{\left( 2 \pi \right)^{3}}
\left[
-  \left( Z_1^A \left( \bk \right) + S_1^A \left( \bk \right) \right)  \left( Z_3^B \left(\bk ,  \bq ,-\bq \right)  +S_3^B \left(\bk ,  \bq ,-\bq \right)  \right)
\right.
\nonumber \\
&& \left. \qquad
+  \left( Z_1^B \left( \bk \right)+ S_1^B \left( \bk \right) \right) \left( Z_3^A \left(\bk ,  \bq ,-\bq \right) +S_3^A \left(\bk ,  \bq ,-\bq \right) \right)
 \right]
   P(q) \, .
\eea

At this point we derive the dipole of the power spectrum, through the definition Eq.~\eqref{def_multipole}. 
At leading order we have
\be
P_1^{(11)} \left( k \right) =
\left[ - i \Delta b_1 \mathcal{R} f  \frac{\HH}{k} 
- 
\sigma^2  \frac{\HH}{k} i f {\mathcal{R}}  \left({ \frac{22}{189}\Delta b_1 -\frac{10 \Delta b_2 }{7}-\frac{\Delta b_3}{2} } \right) 
\right]
P\left( k,z \right) \, .
\ee
The next-to-leading order dipole contributions are given as follow
\bea
\label{P22dip}
P^{(22)}_1 \left( k \right)  
&=&
 \Omega_M \frac{\HH}{k} i  \int \frac{d^3 q }{\left( 2 \pi \right)^3} \left[\sum_{\Delta b= \left\{ \Delta b_1,\Delta b_2,\Delta b_{12} \right\}} \Delta b  \ J_{22}^{\Delta b} \left( \frac{q}{k},\hat \bk \cdot \hat \bq \right) \right] P \left( q \right) P \left( \left| \bk - \bq \right| \right)
  \nonumber \\
  &&
  + \frac{\HH}{k}  i \int \frac{d^3 q }{\left( 2 \pi \right)^3} \left[\sum_{\Delta b= \left\{ \Delta b_1,\Delta b_2,\Delta b_{12} \right\}} \Delta b  \ I_{22}^{\Delta b} \left( \frac{q}{k},\hat \bk \cdot \hat \bq \right) \right]  P \left( q \right) P \left( \left| \bk - \bq \right| \right)
  \\
  P^{(13)}_1 \left( k \right) &=&
 \Omega_M \frac{\HH}{k}   i P \left( k \right) \int \frac{d^3 q }{\left( 2 \pi \right)^3} \left[
 \sum_{\Delta b= \left\{ \Delta b_1,\Delta b_2,\Delta b_{12} \right\}} \Delta b  \ J_{13}^{\Delta b} \left( \frac{q}{k}\right)
 \right] P \left( q \right) 
  \nonumber \\
  &&
  + \frac{\HH}{k} i  P \left( k \right) \Delta b_1 \int \frac{d^3 q }{\left( 2 \pi \right)^3} I_{13}^{\Delta b_1} \left( \frac{q}{k} \right)  P \left( q \right) 
  \nonumber \\
  &&
  + \frac{\HH}{k} i P \left( k \right) \sigma^2 \left[
 \sum_{\Delta b= \left\{ \Delta b_1,\Delta b_2, \Delta b_3 \right\}} \Delta b  \ K_{13}^{\Delta b} 
 \right]
 \label{P13dip}
\eea
where $J_{22}^{\Delta b},I_{22}^{\Delta b},J_{13}^{\Delta b},I_{13}^{\Delta b}$ are defined in Appendix~\ref{app:dipole_kernels}. We have also separated the contribution induced by gravitational potential terms, where the matter density parameter $\Omega_M$ appears through the Poisson equation. We remark that the last line of Eq.~\eqref{P13dip} vanishes together with the last line of Eq.~\eqref{P11dip}, leading to a total finite result.

\section{Numerical results}
\label{sec:numres}
We evaluate numerically the dipole of the correlation function computed in the previous section. We use the following cosmology\footnote{In order to compare, whenever it is possible with Ref.~\cite{Breton:2018wzk}, we adopt the same cosmology.}: $h=0.72$, $\Omega_{\rm M} =0.25733 $, $\Omega_{\rm b} = 0.04356$. The primordial curvature power spectrum is determined by $\sigma_8 =0.801 $, the spectral index $n_s  = 0.963$ and no running. The matter power spectrum is computed with \class{}~\cite{Blas:2011rf,DiDio:2013bqa}.

 \begin{figure}[h!]
\begin{center}
\includegraphics[width=0.32\textwidth]{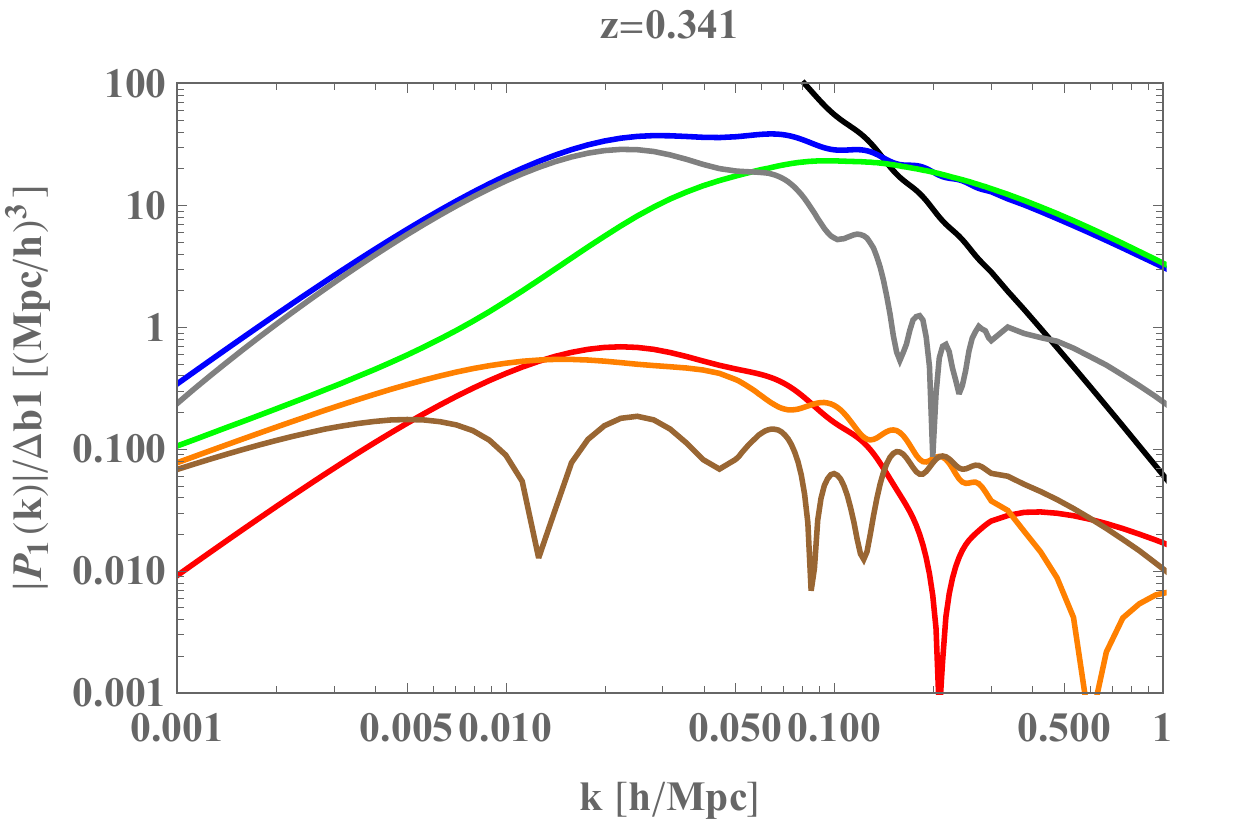} 
\includegraphics[width=0.32\textwidth]{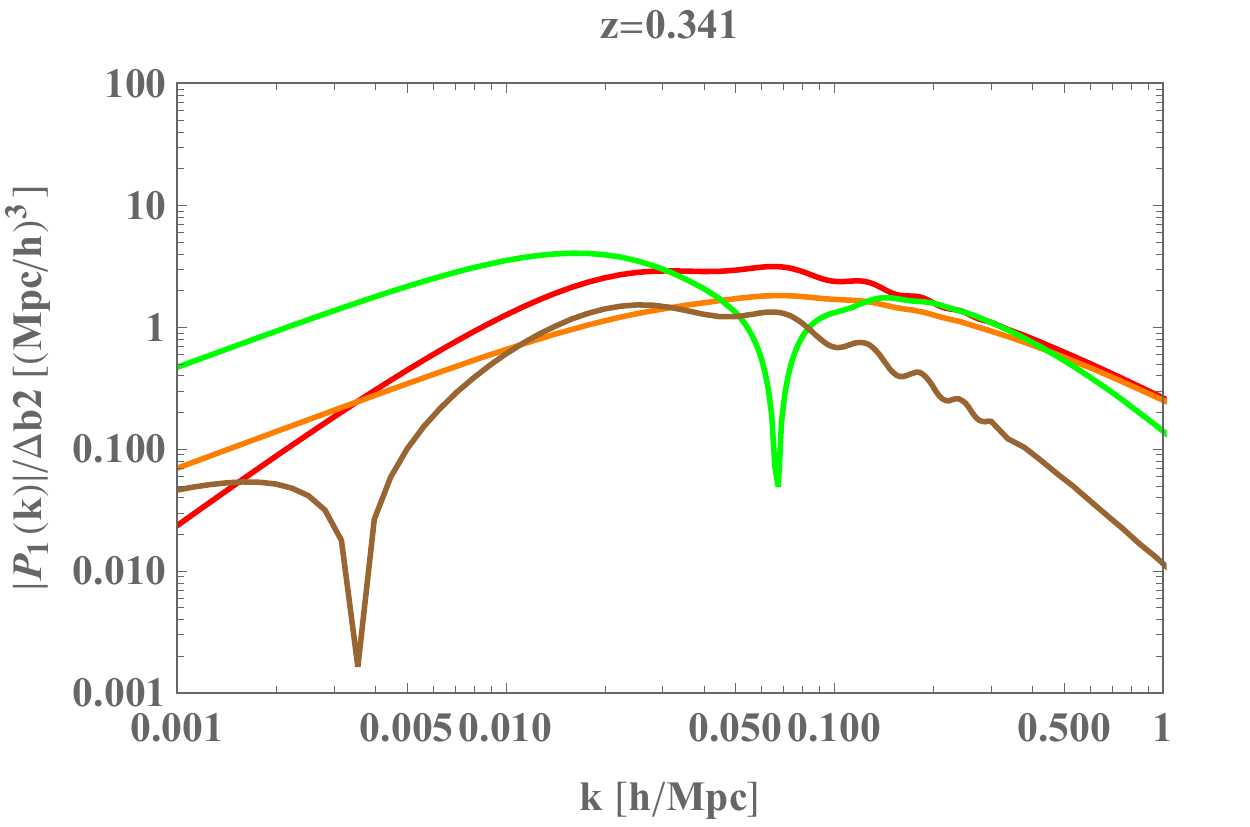}  
\includegraphics[width=0.32\textwidth]{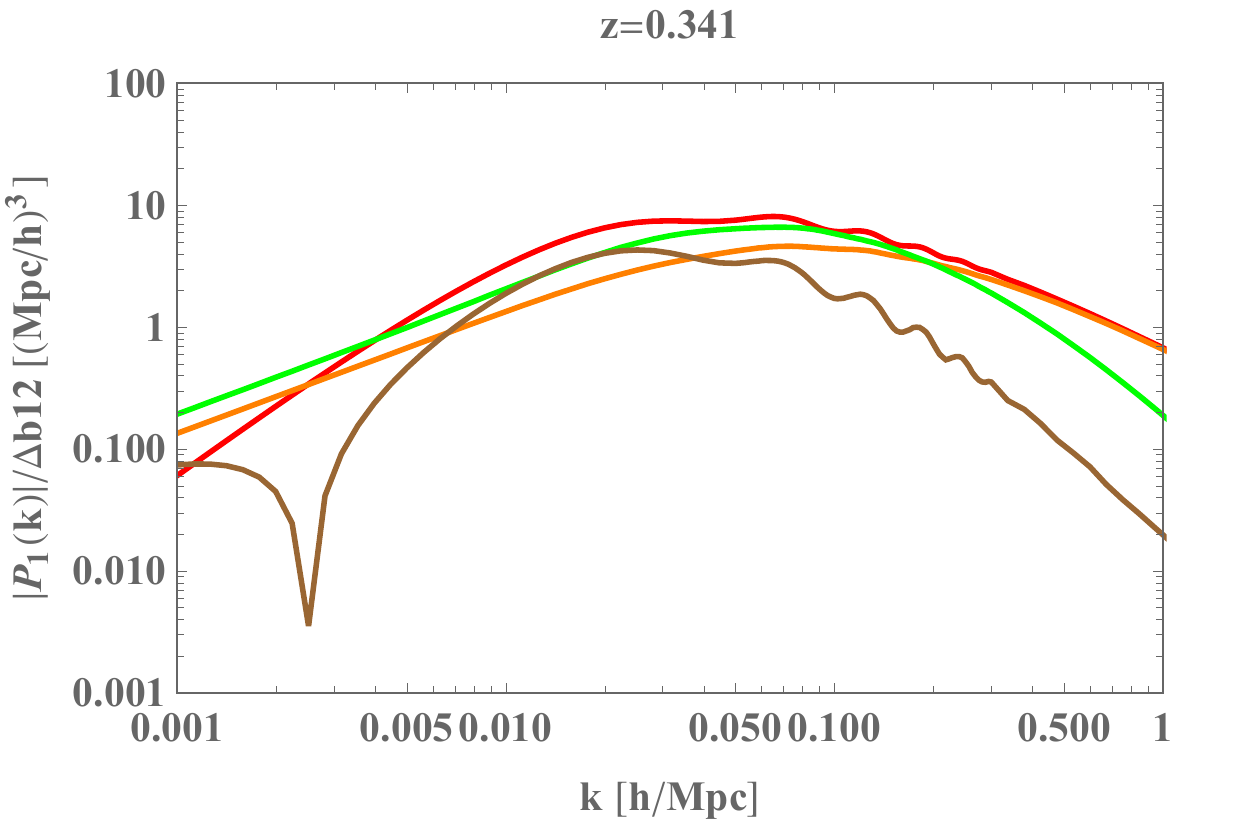}\\
\includegraphics[width=0.32\textwidth]{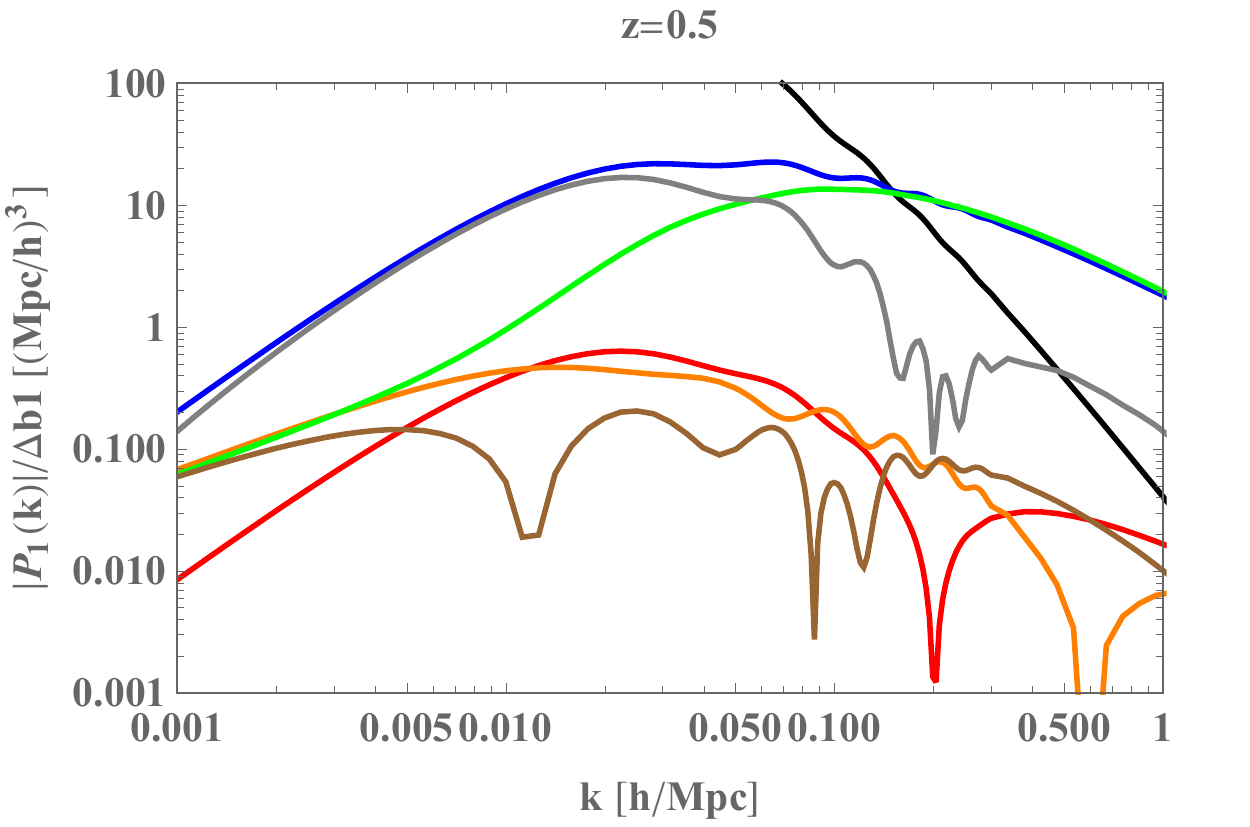} 
\includegraphics[width=0.32\textwidth]{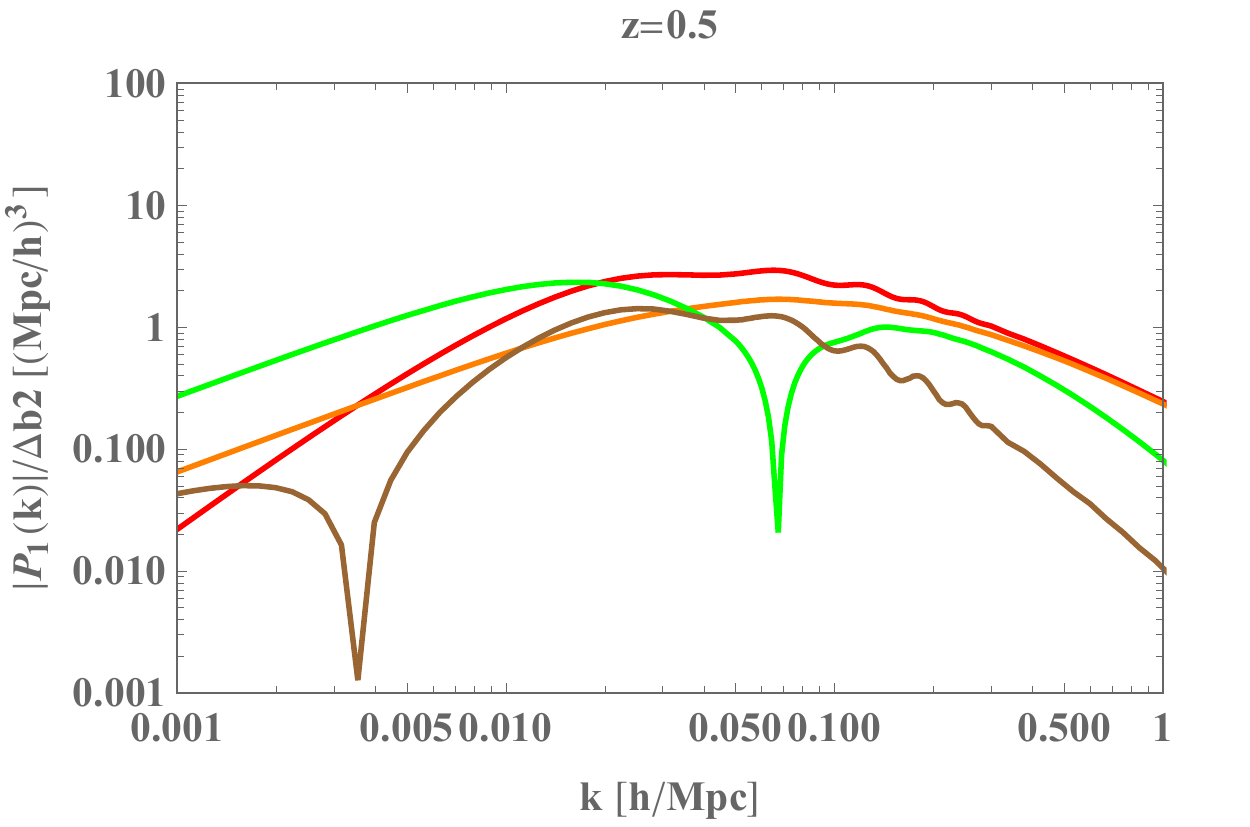}  
\includegraphics[width=0.32\textwidth]{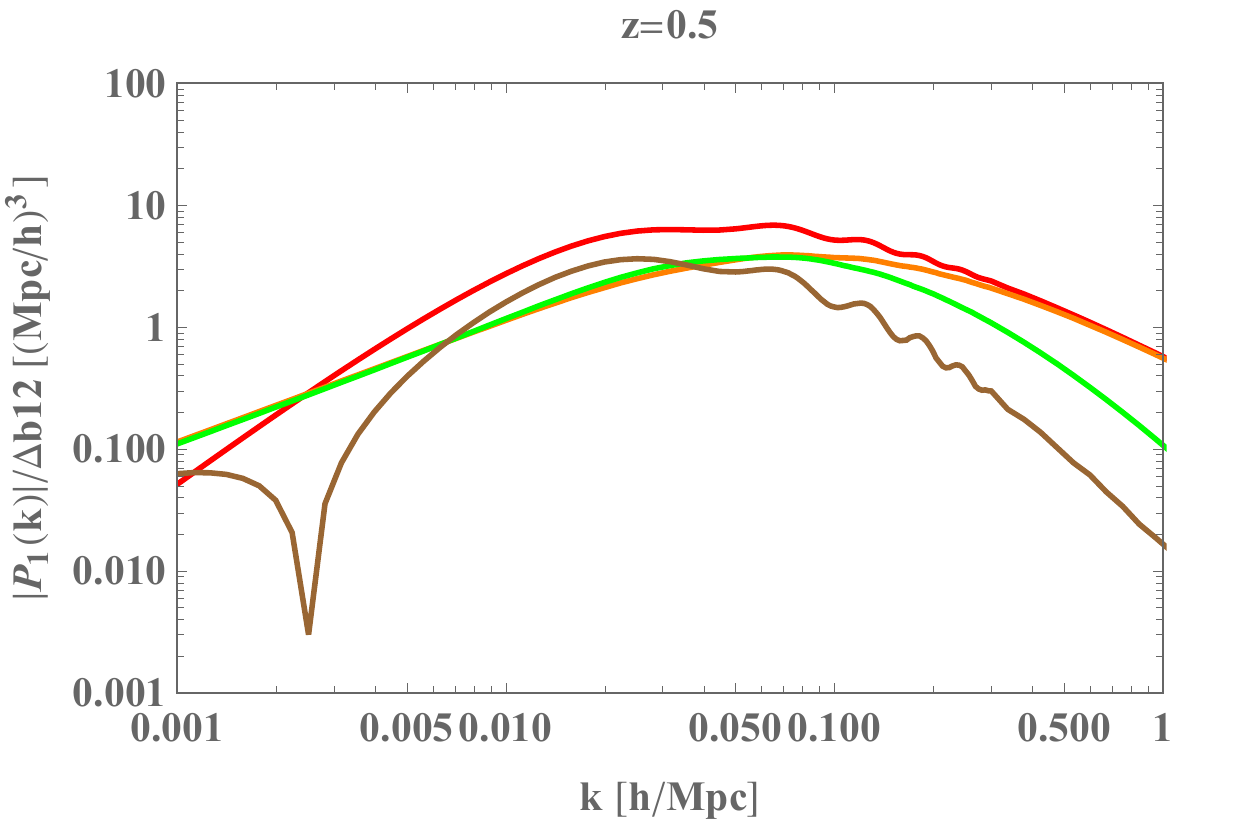}\\
\includegraphics[width=0.32\textwidth]{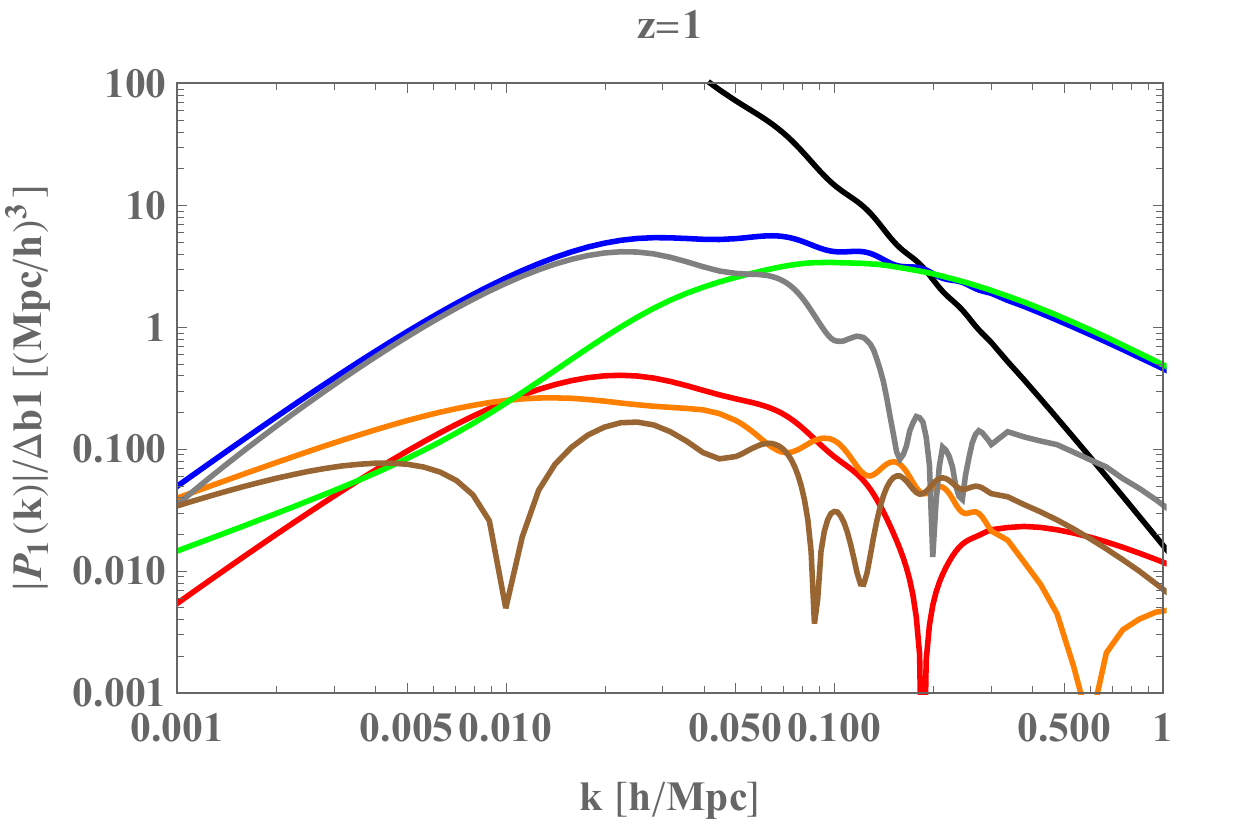} 
\includegraphics[width=0.32\textwidth]{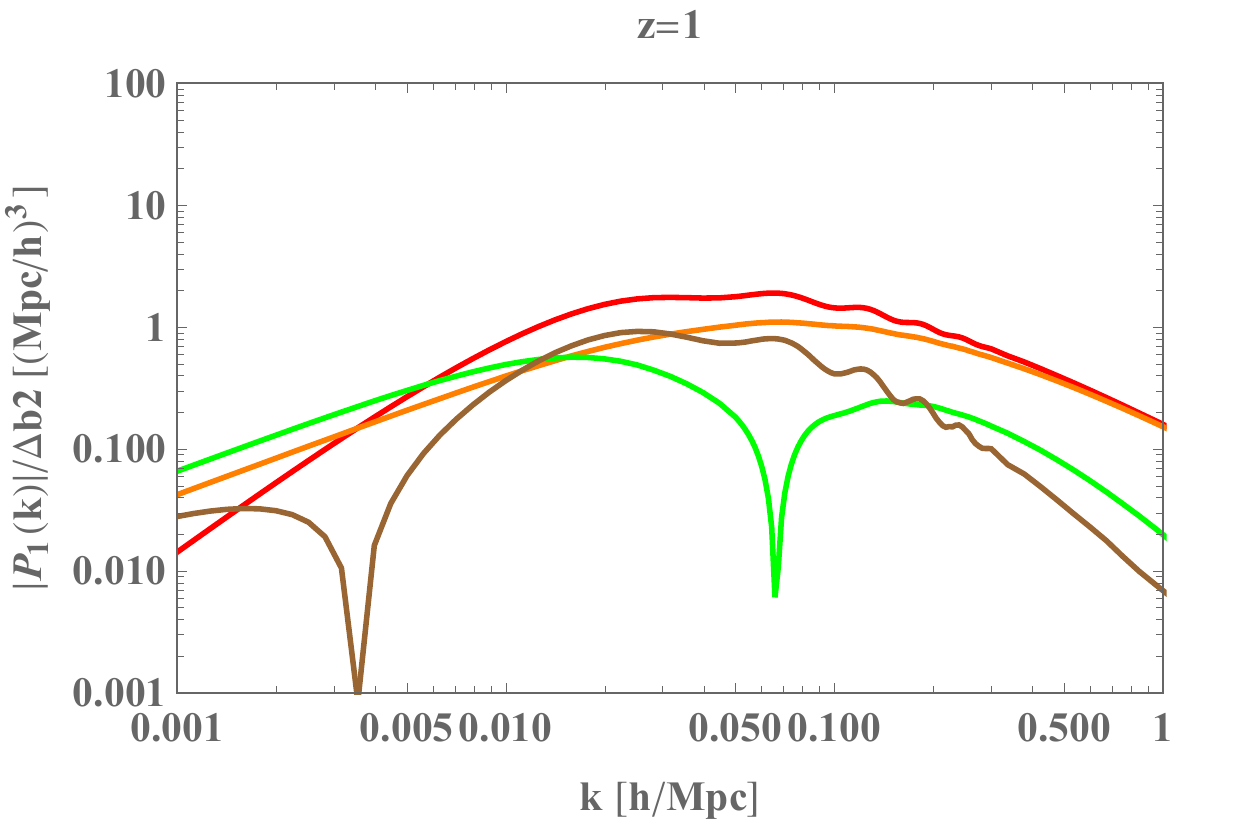}  
\includegraphics[width=0.32\textwidth]{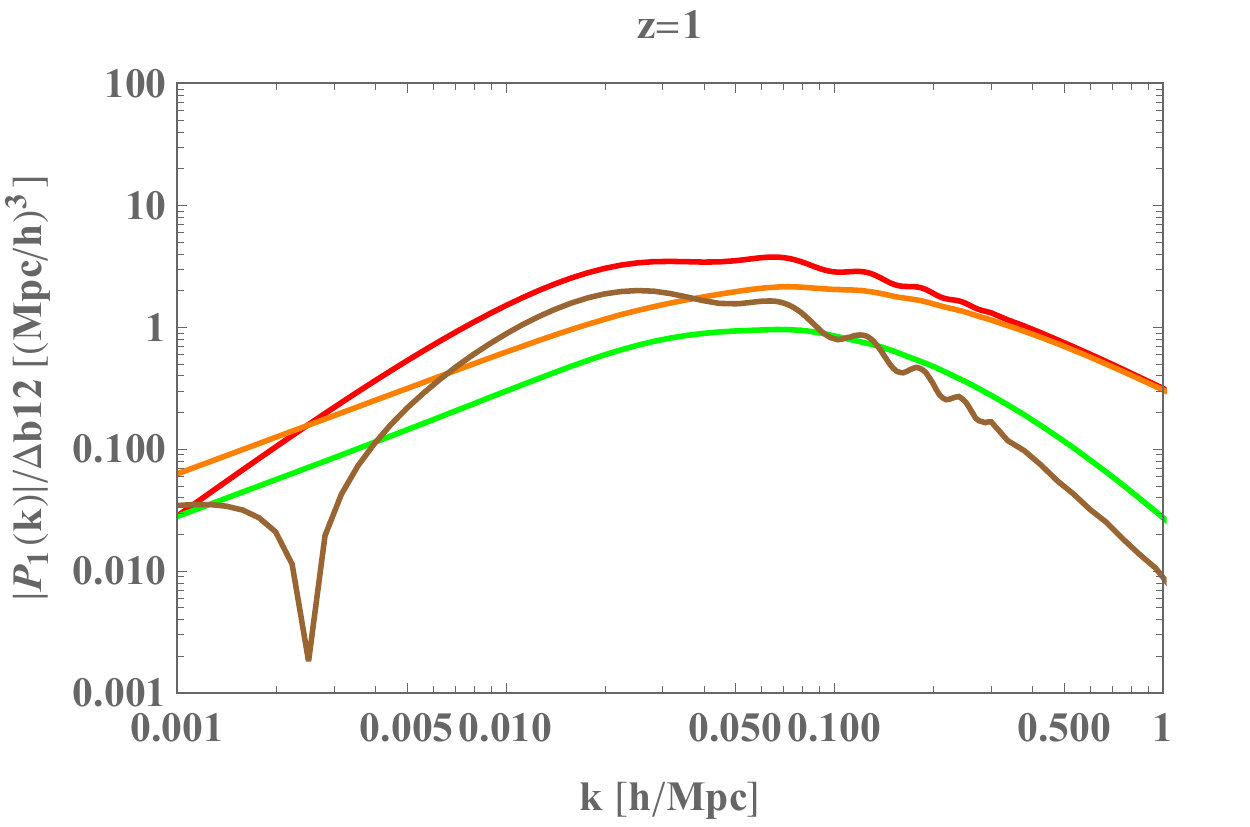}
\caption{We plot the absolute value of the dipole of the power spectrum at three different redshifts: $z=0.341$ (top), $z=0.5$ (center), $z=1$ (bottom). We show separately the terms proportional to $\Delta b_1$ (left), $\Delta b_2$ (center) and $\Delta b_{12}$ (right). Different colours refers to: linear dipole (black), $P^{13}_1$ (red) and $P^{22}_1$ (orange) which depends on the gravitational potential and their sum (brown),  $P^{13}_1$ (blue) and $P^{22}_1$ (green), which do not depend on the gravitational potential and their sum (gray).
}
 \label{fig_powerspectrum}
\end{center}
\end{figure}

To be as general as possible we plot separately the terms involving the different bias coefficients.
From Fig.~\ref{fig_powerspectrum} we notice that the contribution proportional to $\Delta b_1$ of the gravitational potential is always subdominant. This is due to Equivalence Principle, as via
the Euler equation there are no terms of the form ${\bf\grad} \Psi$. The gravitational potential enters only coupled to density or velocity perturbations. Several of the terms do not lead to any dipole because they are even in the bias parameters.
We also notice that amplitudes of different 1-loop corrections are comparable to linear theory at scale about $k \sim  { 0.15} h/{\rm Mpc} $ at $z\sim0.3$ and $k \sim 0.2 h/{\rm Mpc} $ at $z\sim1$, as expected since this is 
where power per mode is roughly unity. 
We clearly see that there are strong cancellations between $P_1^{13}$ and $P_1^{22}$ which suppress the 1-loop corrections. This is expected and indeed 
acts as a check of our relativistic derivation to third order in perturbation theory.

It is interesting to observe the redshift dependence of the next-to-leading order correction with respect to the linear prediction. While we do expect non-linearity to grow with time, the pre-factor $(\HH r)^{-1}$ of the linear dipole is becoming larger at low redshift. Because of that, the dipole is less affected by 1-loop correction at late time when compared to even multipoles. The amplitude of 1-loop corrections are only weakly redshift dependent.
Consequently also the gravitational potential terms become more relevant (with respect to the terms dominated by peculiar velocities) at higher redshifts.
 \begin{figure}[h!]
\begin{center}
\includegraphics[width=0.32\textwidth]{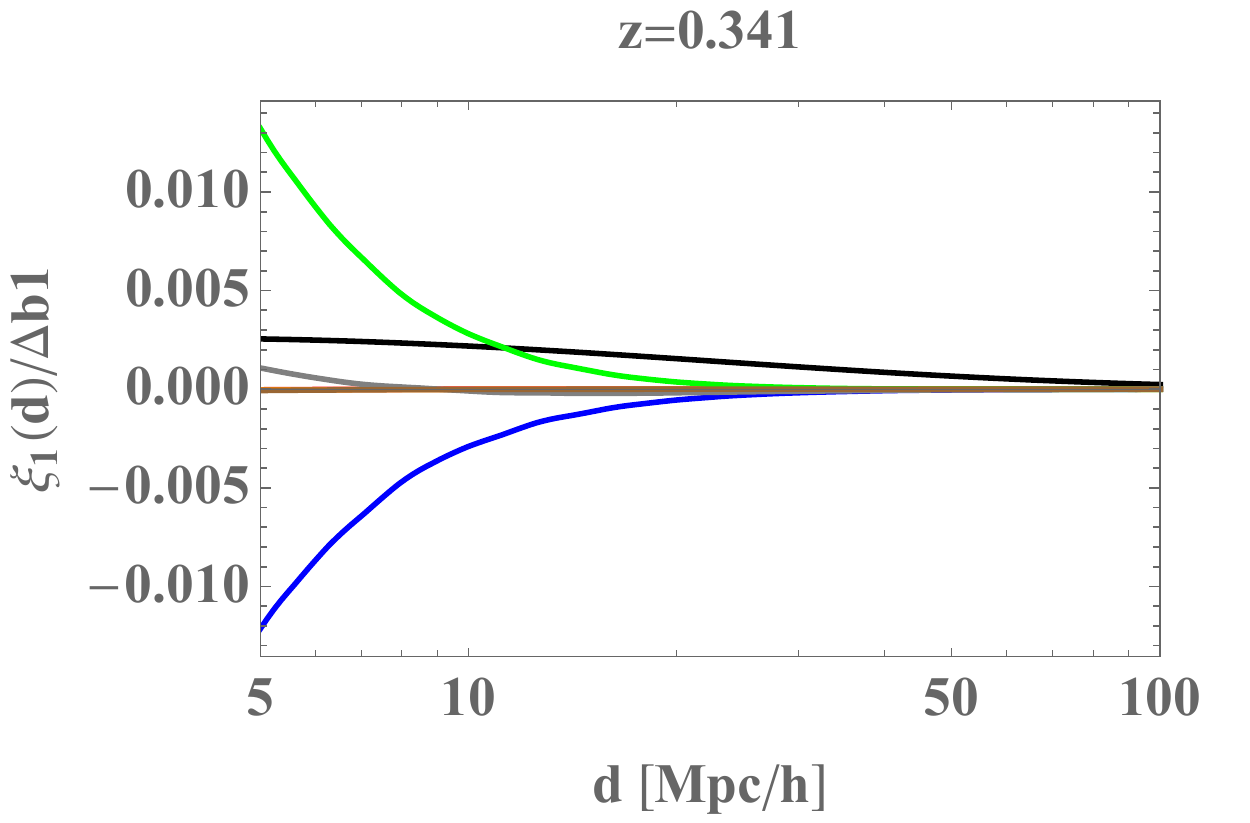} 
\includegraphics[width=0.32\textwidth]{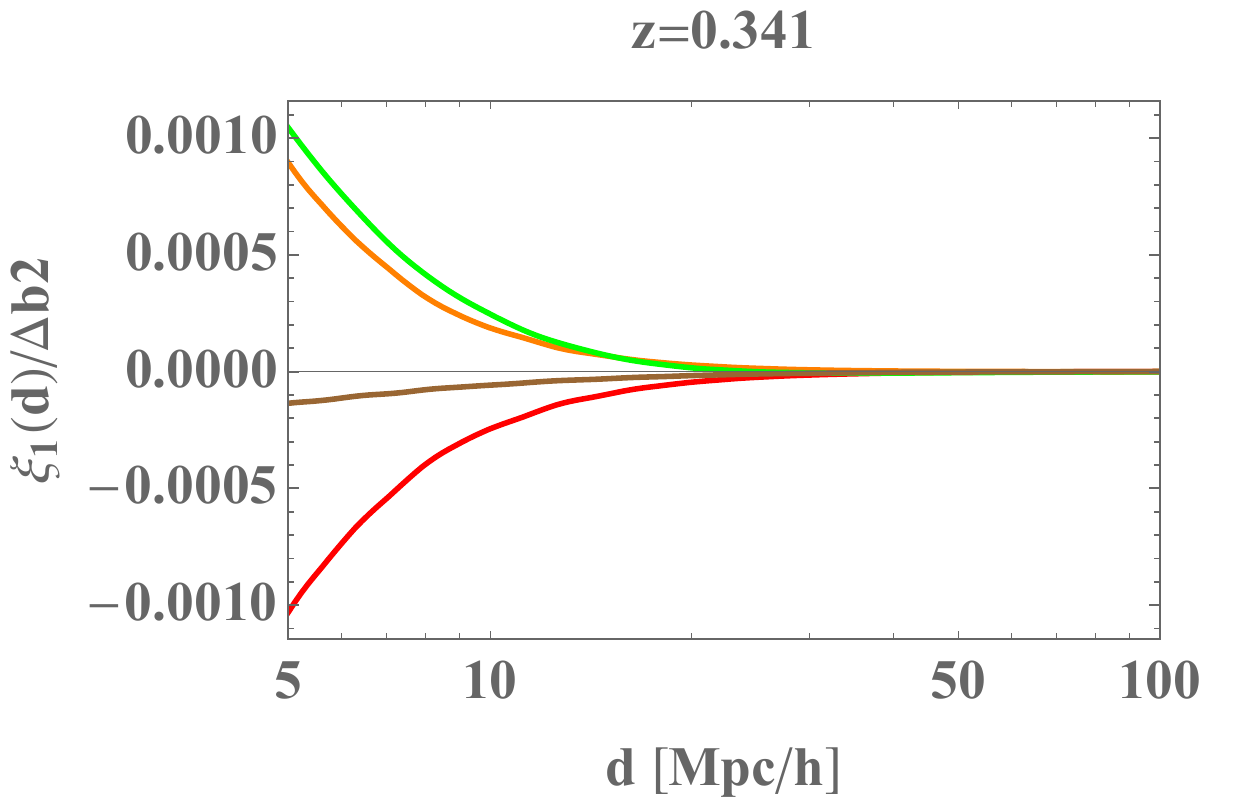}  
\includegraphics[width=0.32\textwidth]{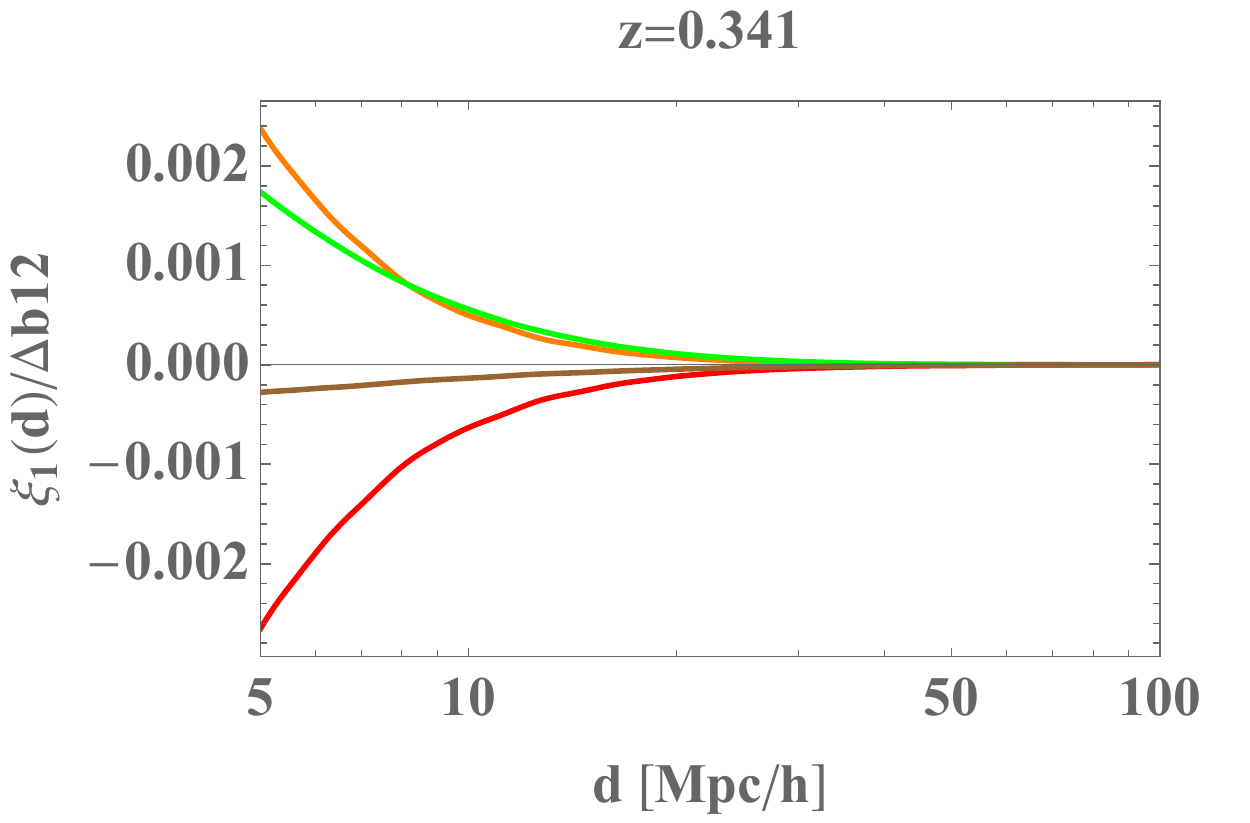}\\
\includegraphics[width=0.32\textwidth]{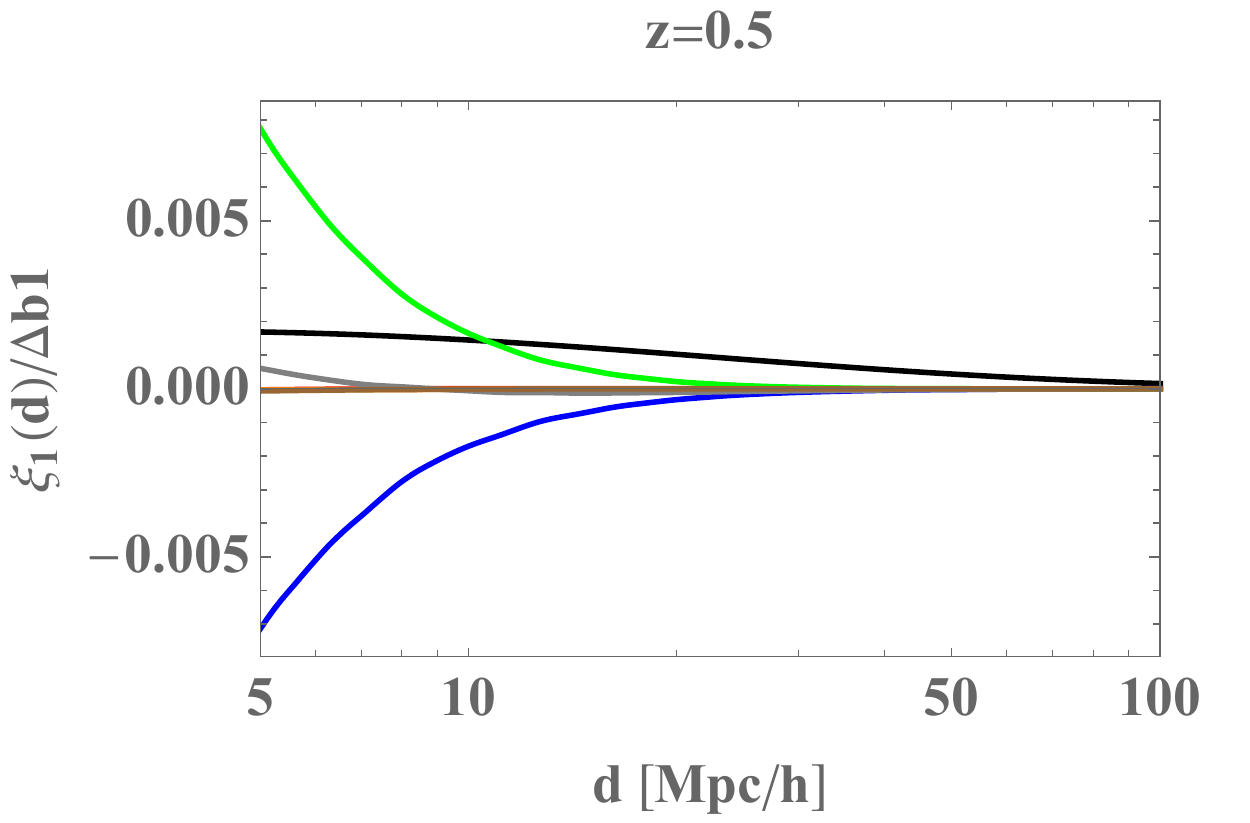} 
\includegraphics[width=0.32\textwidth]{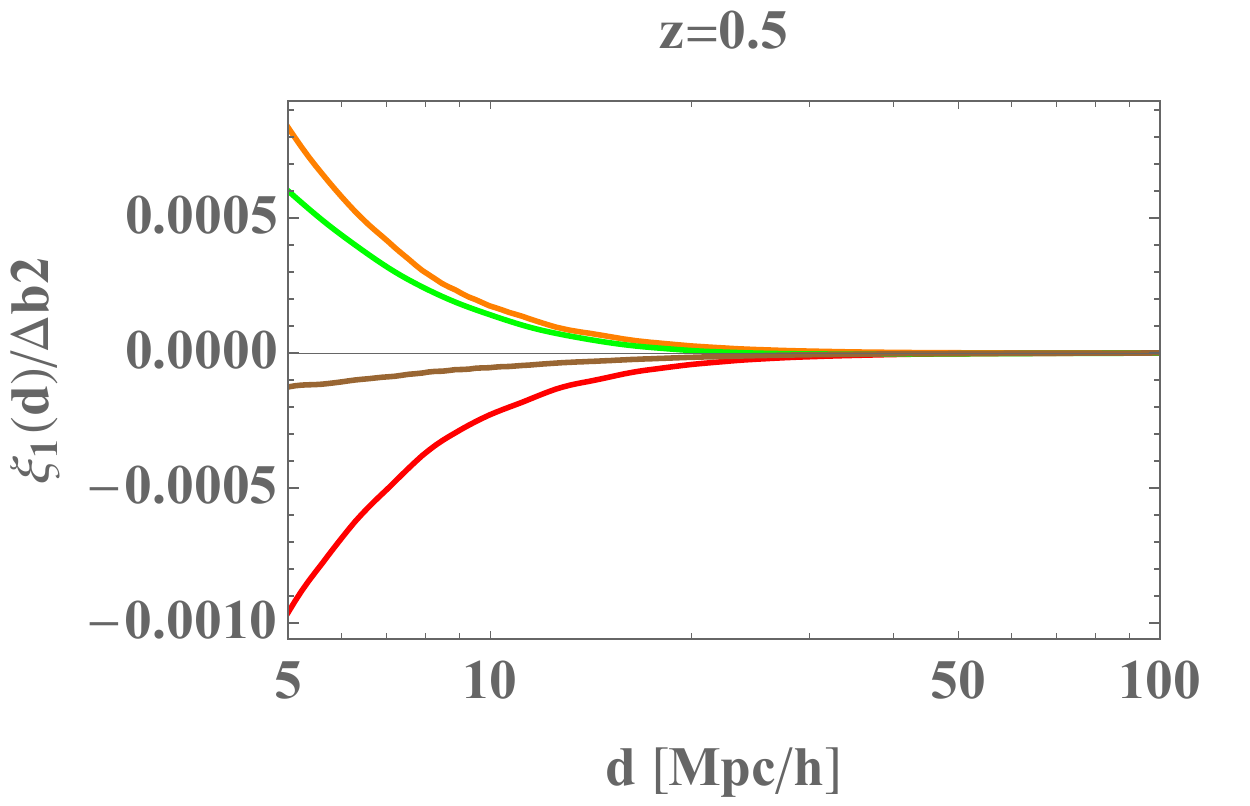}  
\includegraphics[width=0.32\textwidth]{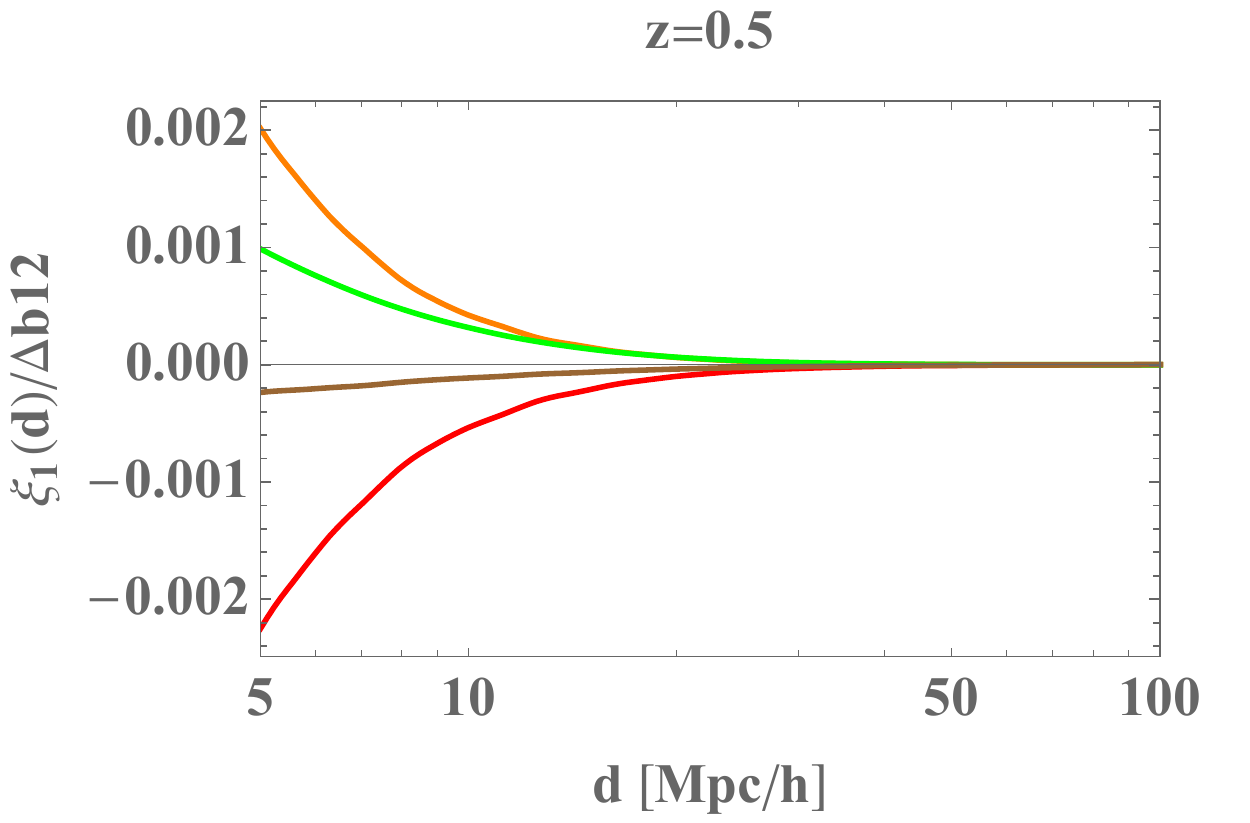}\\
\includegraphics[width=0.32\textwidth]{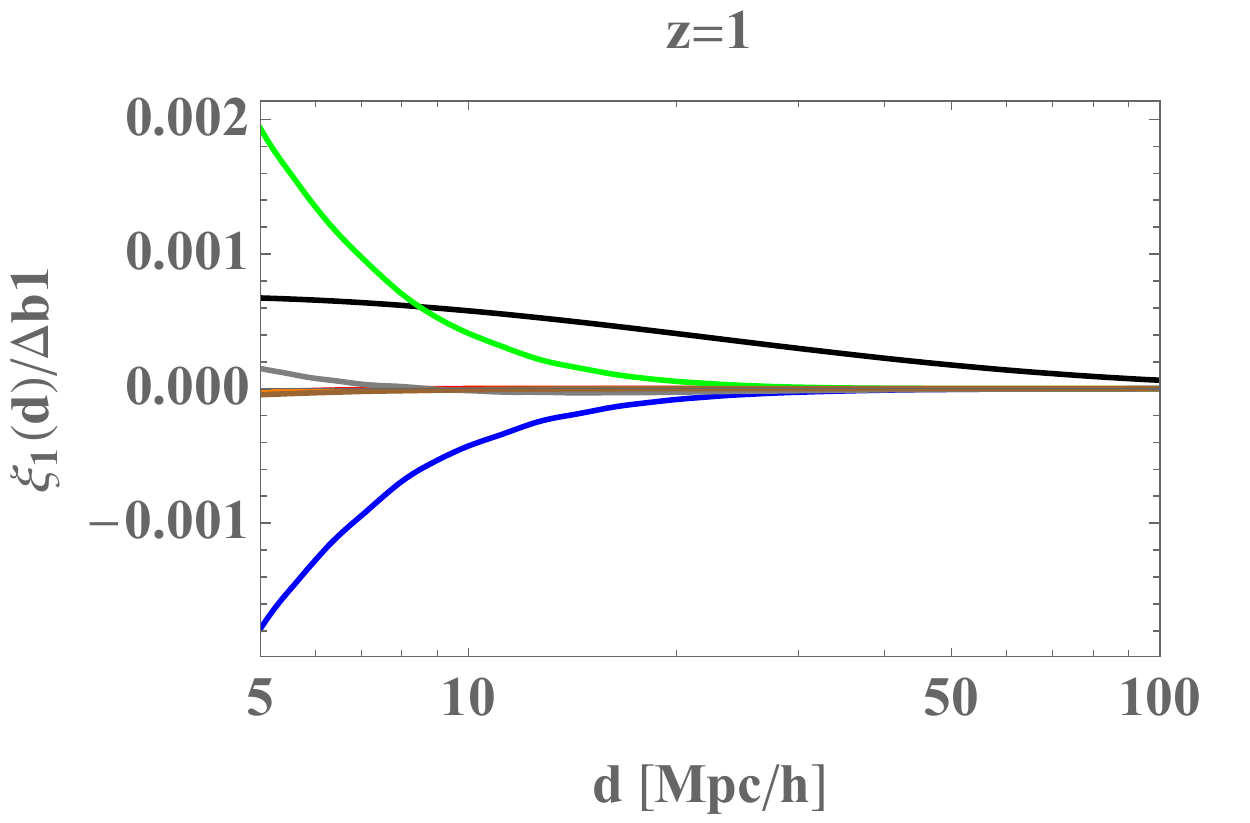} 
\includegraphics[width=0.32\textwidth]{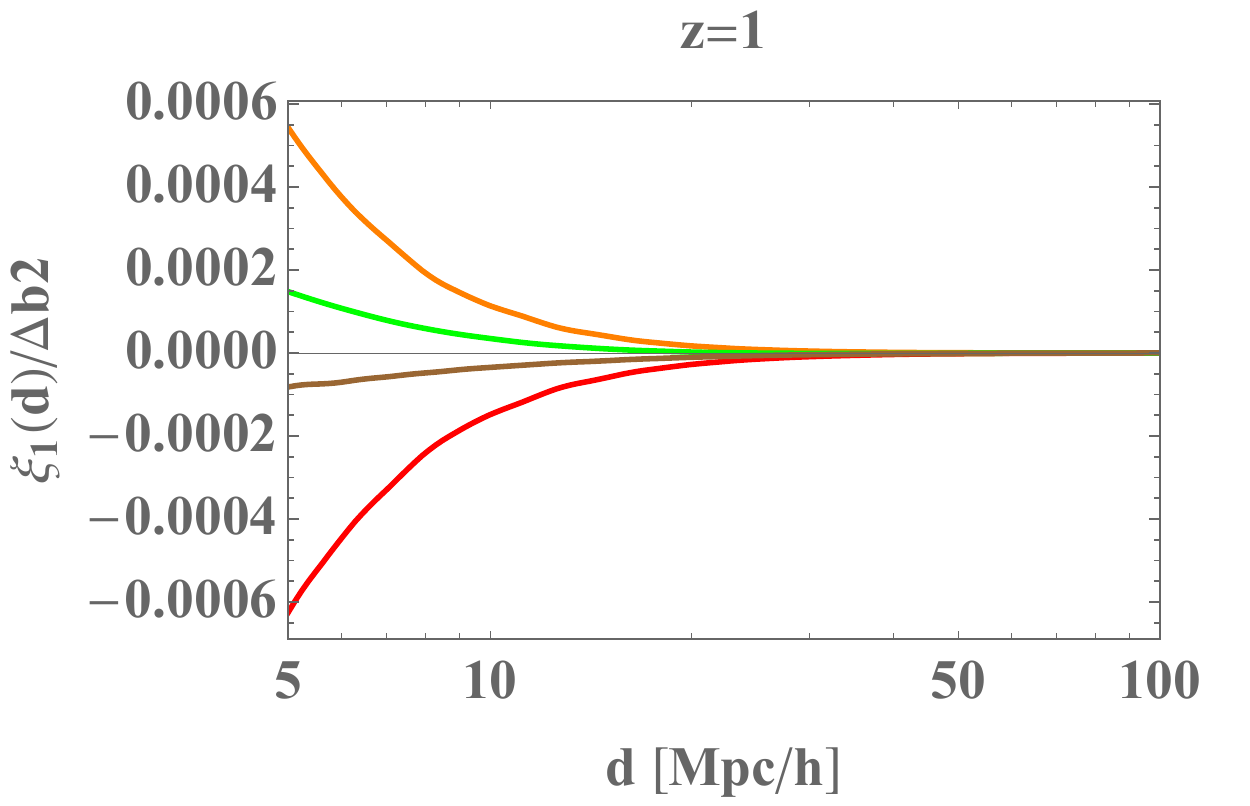}  
\includegraphics[width=0.32\textwidth]{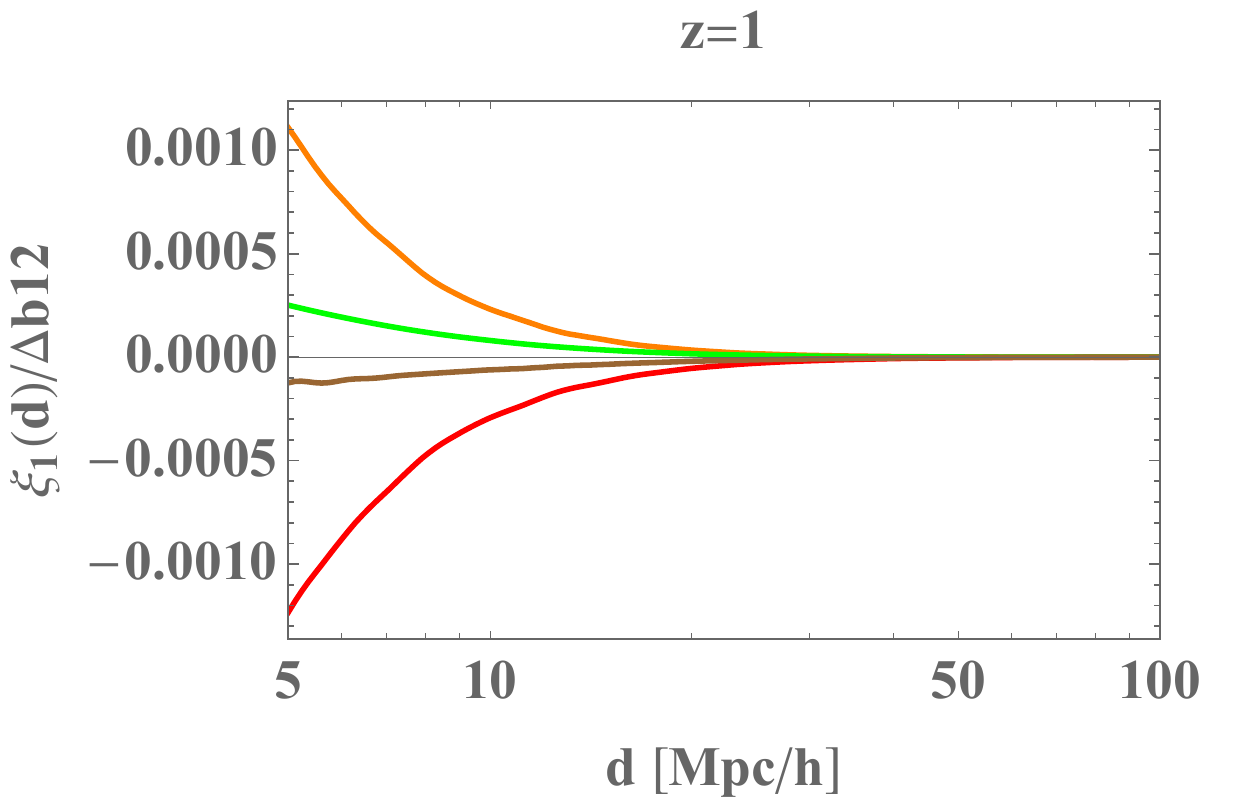}
\caption{We plot the dipole of the correlation function at three different redshifts: $z=0.341$ (top), $z=0.5$ (center), $z=1$ (bottom). We show separately the terms proportional to $\Delta b_1$ (left), $\Delta b_2$ (center) and $\Delta b_{12}$ (right). We adopt the same colors scheme of Fig.~\ref{fig_powerspectrum}.
}
 \label{fig_correlation}
\end{center}
\end{figure}

\begin{figure}[h!]
\begin{center}
\includegraphics[width=0.32\textwidth]{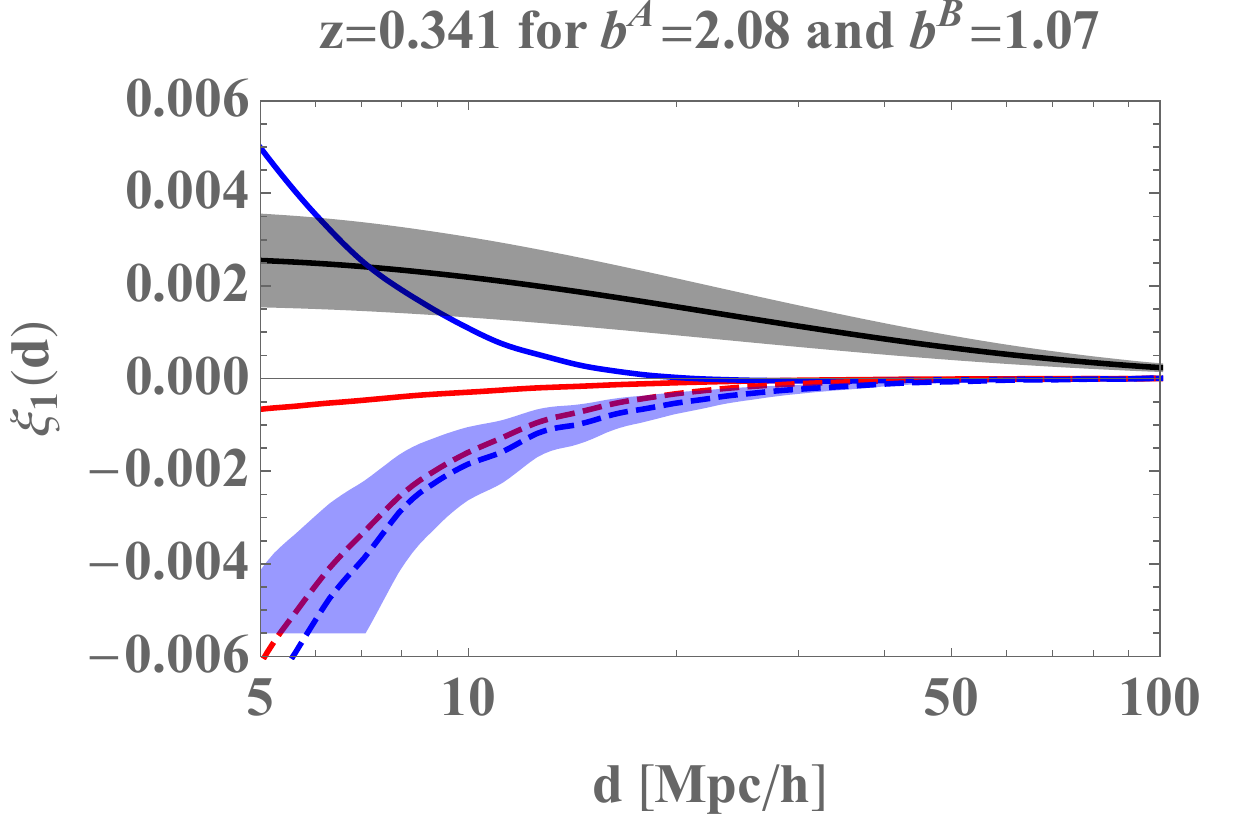} 
\includegraphics[width=0.32\textwidth]{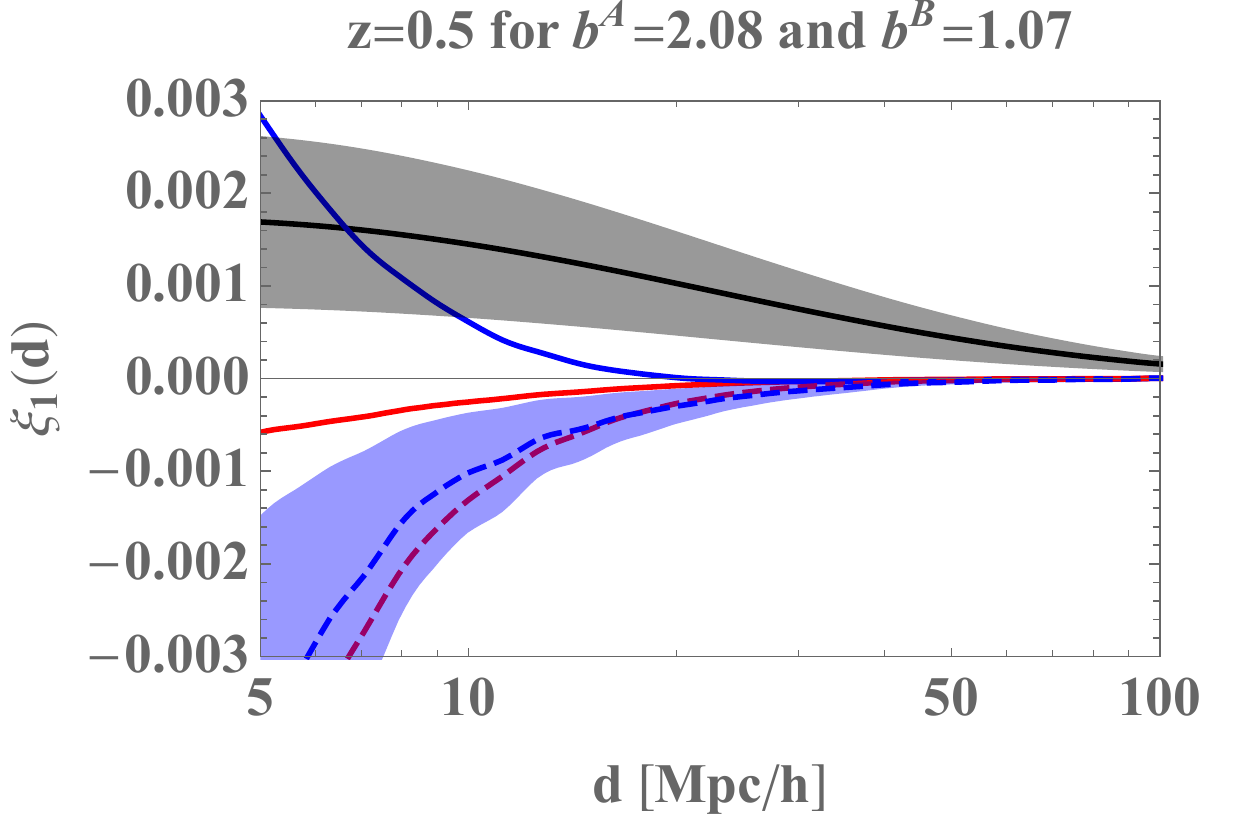}  
\includegraphics[width=0.32\textwidth]{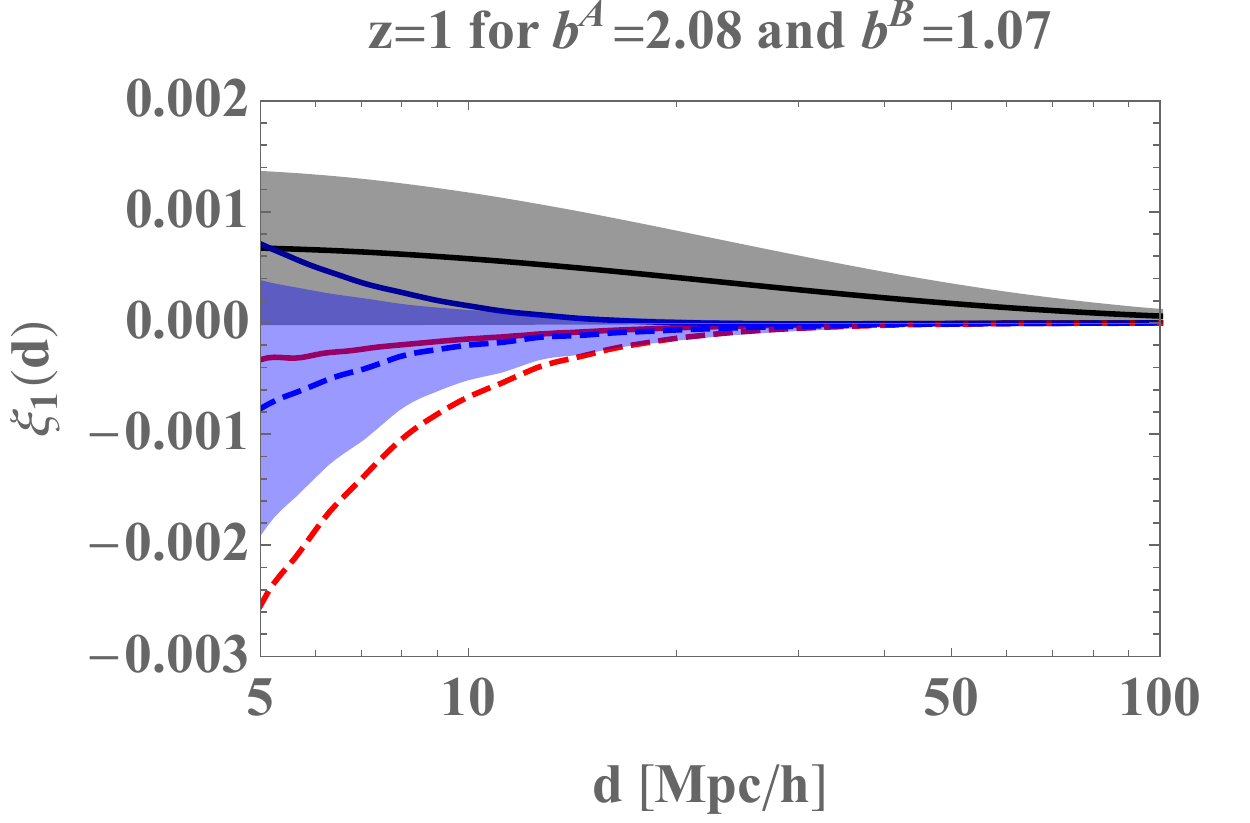}
\caption{We plot the dipole of the correlation function at three different redshifts: $z=0.341$ (left), $z=0.5$ (center), $z=1$ (right) for the local bias parameters $b_1^A = 2.08$ and  $b_1^B = 1.07$, by adopting Eq.~\eqref{bias_relation}. Black line denotes the linear dipole, red includes all the terms involving gravitational potential, and blue all the other terms. Dashed (solid) lines include (neglect) the non-linear velocity dispersion parameter, see Appendix~\ref{app:veldisp}. The black and blue regions represent the dipole for different values of $f_{\rm evo} \in \left[ -3, 3 \right].$
} \label{fig_correlation_tot}
\end{center}
\end{figure}

By using Eq.~\eqref{xi_ell} we can show, see Fig.~\ref{fig_correlation}, the 1-loop corrections also to the dipole of the correlation function.
Through the mass-dependence of the local halo-bias parameters it is possible to rewrite $b_2$ as a function of $b_1$. We adopt the numerical fit provided by Ref.~\cite{Lazeyras:2015lgp}
\be \label{bias_relation}
b_2\left( b_1 \right) = 0.412 - 2.143 b_1+0.929 b_1^2 +0.008 b_1^3 \, .
\ee
Therefore we can now plot, see Fig.~\ref{fig_correlation_tot}, the dipole of the correlation function just as a function of the local bias parameters $b_1$ for the two galaxy populations. As pointed out by Ref.~\cite{Breton:2018wzk}, on small scales the dipole does not scale simply as the difference between $b_1^A $ and $b_1^B$, but it depends explicitly on the amplitude of the two bias parameters. As shown in the next section, our perturbative approach recovers the correct dependence on the values of the two biases $b_1^A $ and $b_1^B$ on small scales.
From Fig.~\ref{fig_correlation_tot} we see that the terms induced by peculiar velocity (blue) are severely suppressed when going to higher redshift as compared to the gravitational terms (red). This behavior is caused by the presence of the prefactor $ \left( \HH r \right)^{-1}$ which dominates the peculiar velocity terms. This indicates that galaxy samples at higher redshift are more suitable for searching of detection of the gravitational redshift effect on LSS, if the linear dipole can be properly subtracted.
The terms induced by the gravitational potentials are purely geometrical and independent from any values of evolution or magnification biases. In fig.~\ref{fig_correlation_tot} we show how the amplitudes of the dipole induced by velocity effects depend on evolution bias $f_{\rm evo} \in \left[ -3, 3\right]$. We see that around $10$ Mpc/h the amplitude of the relativistic dipole induced by peculiar velocity is comparable to the effect induced by gravitational redshift. Therefore, any attempt of detecting gravitational redshift on LSS scales needs to consider the impact of relativistic effects and requires as well an accurate knowledge of evolution and magnification biases.

\section{Comparison to previous works}
\label{sec:discussion}

\begin{figure}[h!]
\begin{center}
\includegraphics[width=0.32\textwidth]{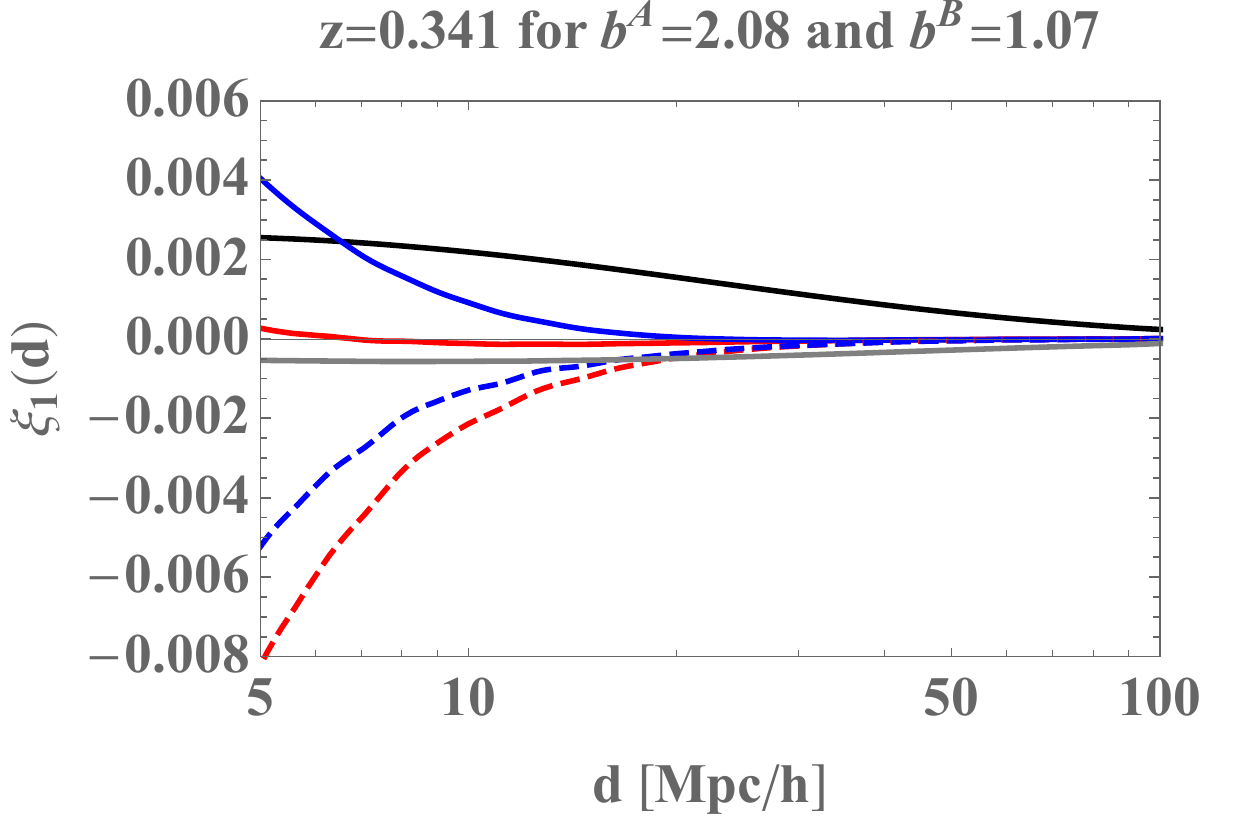} 
\includegraphics[width=0.32\textwidth]{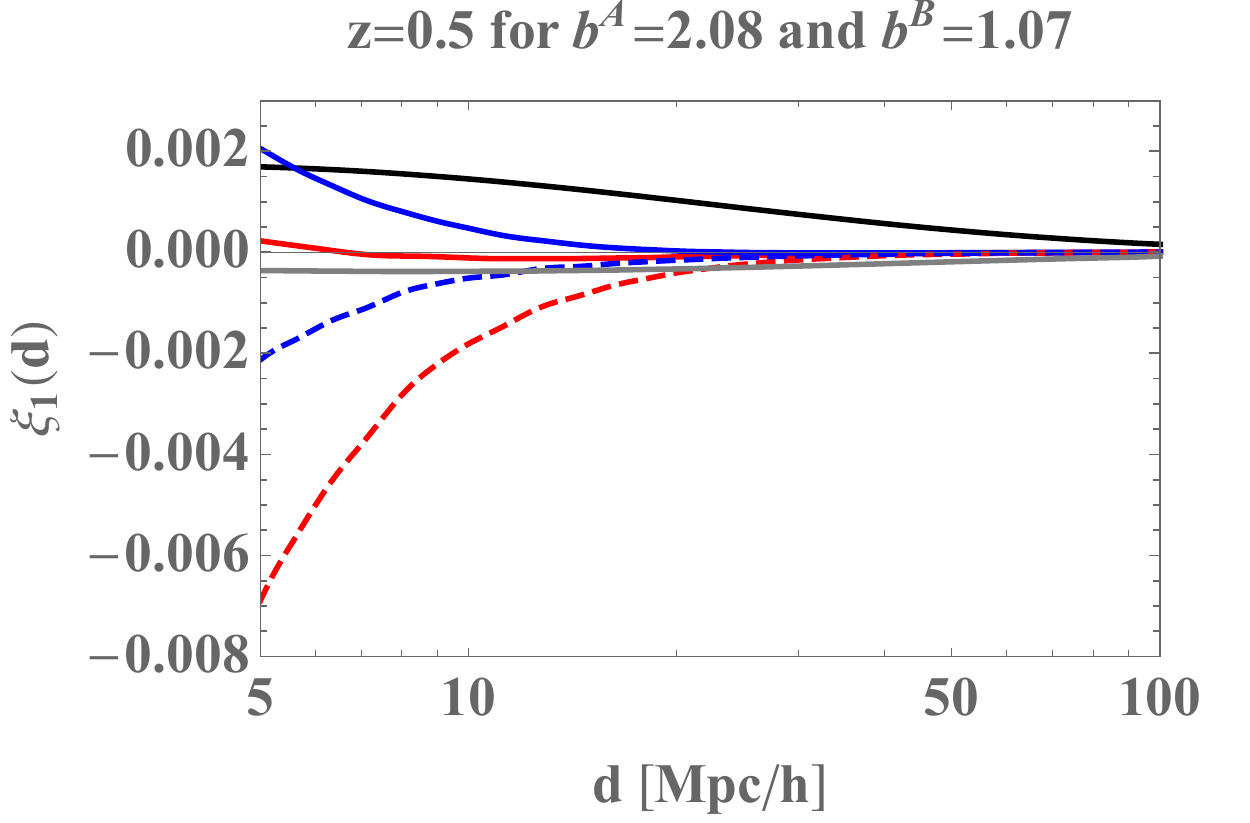}  
\includegraphics[width=0.32\textwidth]{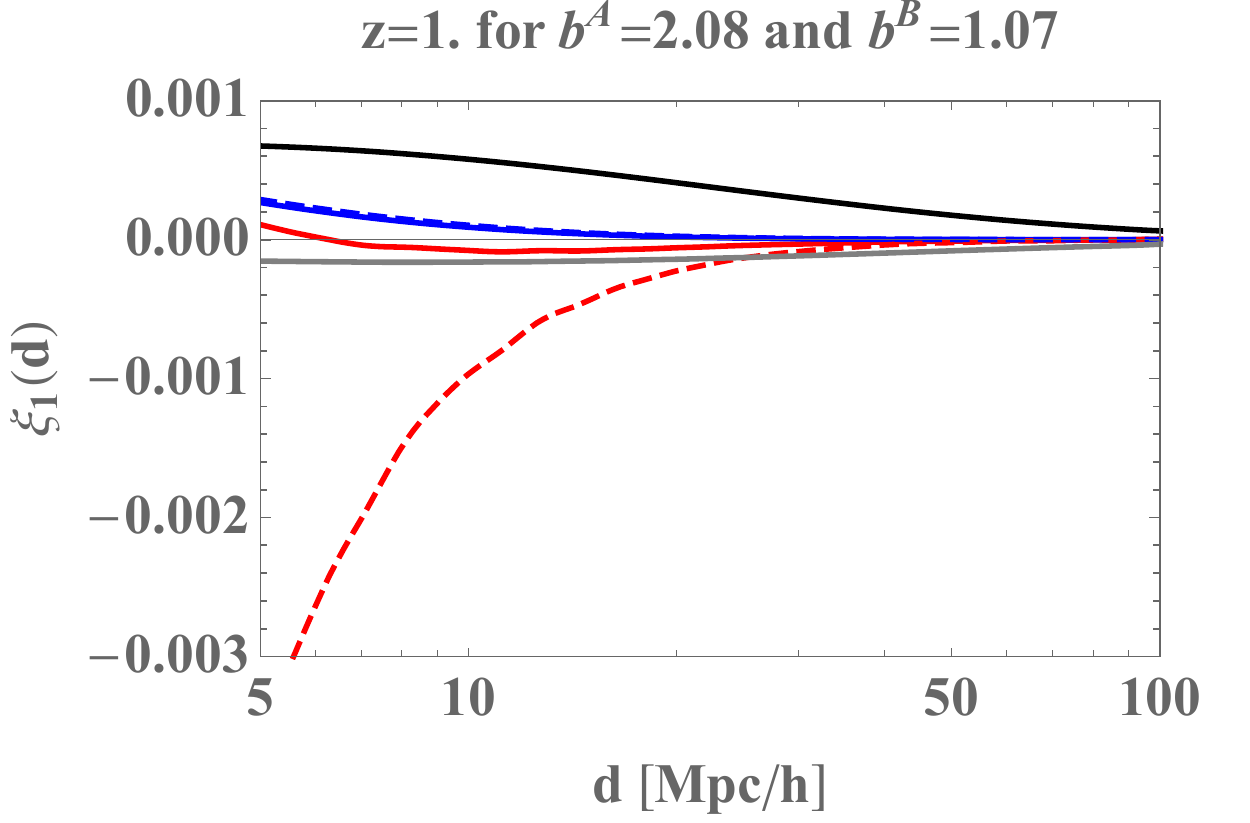}
\caption{We plot the dipole of the correlation function without assuming Euler equation at three different redshifts: $z=0.341$ (left), $z=0.5$ (center), $z=1$ (right) for the local bias parameters $b_1^A = 2.08$ and  $b_1^B = 1.07$, by adopting Eq.~\eqref{bias_relation}. Black line denotes the linear dipole, red includes all the terms involving gravitational potential, and blue all the other terms. Dashed (solid) lines include (neglect) the non-linear velocity dispersion parameter, see Appendix~\ref{app:veldisp}.
The gray line represents the (linear) wide angle effect computed with Eq.~\eqref{eq:WA}. } 
\label{fig_correlation_tot_NoEuler}
\end{center}
\end{figure}

In order to compare with previous results, in particular with Ref.~\cite{Breton:2018wzk}, we recompute the dipole of the correlation function without assuming the Euler equation. Clearly the total dipole will not change, but this allows us to more fairly compare to other works. It is important to remark that while the total dipole does not change on whether we adopt or not the Euler equation, the relative relevance of the terms induced by the gravitational potential does change. Fig.~\ref{fig_correlation_tot_NoEuler} shows the comparison between the different contributions to the dipole without assuming Euler equation. By comparing with fig.~\ref{fig_correlation_tot}, we clearly see that the amplitude of the gravitational redshift is enhanced and the peculiar velocity suppressed. Therefore, not accounting for the Euler equation leads to an overestimation of the amplitude of the gravitational redshift.
\begin{figure}[h!]
\begin{center}
\includegraphics[width=0.45\textwidth]{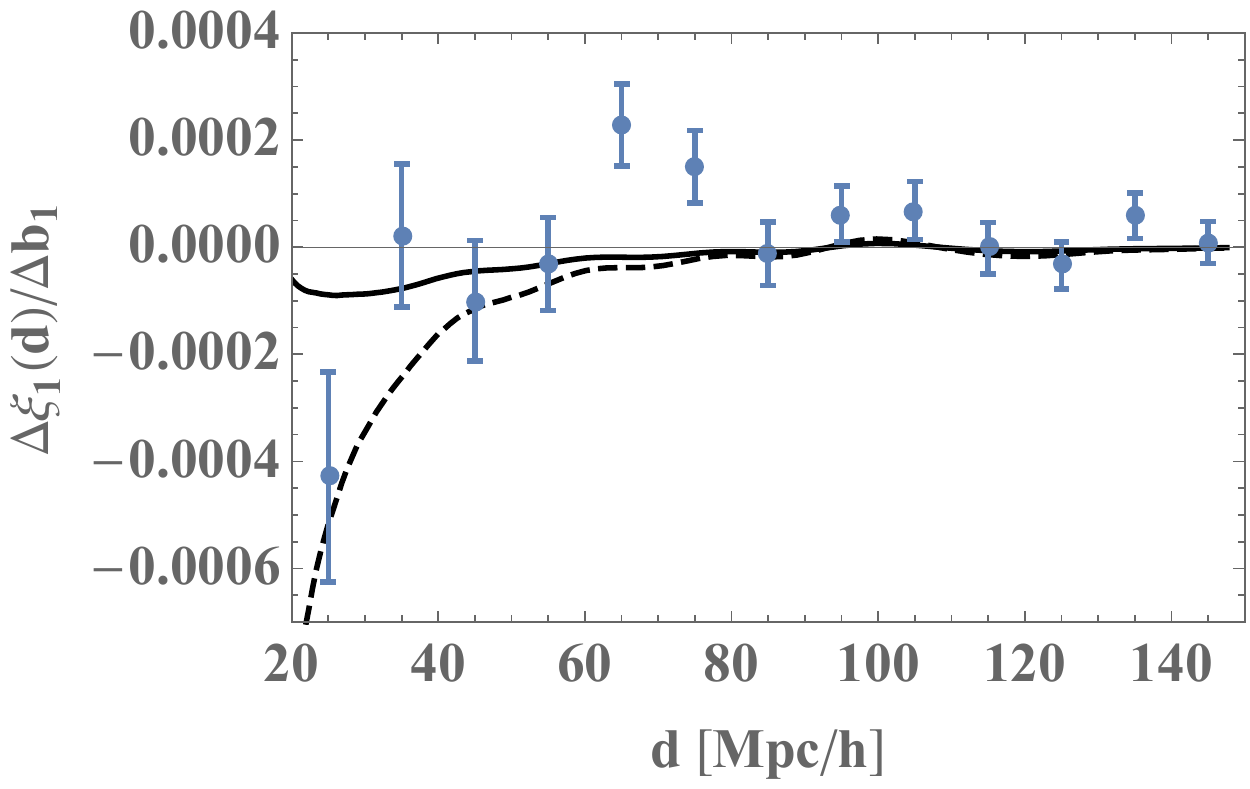} 
\includegraphics[width=0.45\textwidth]{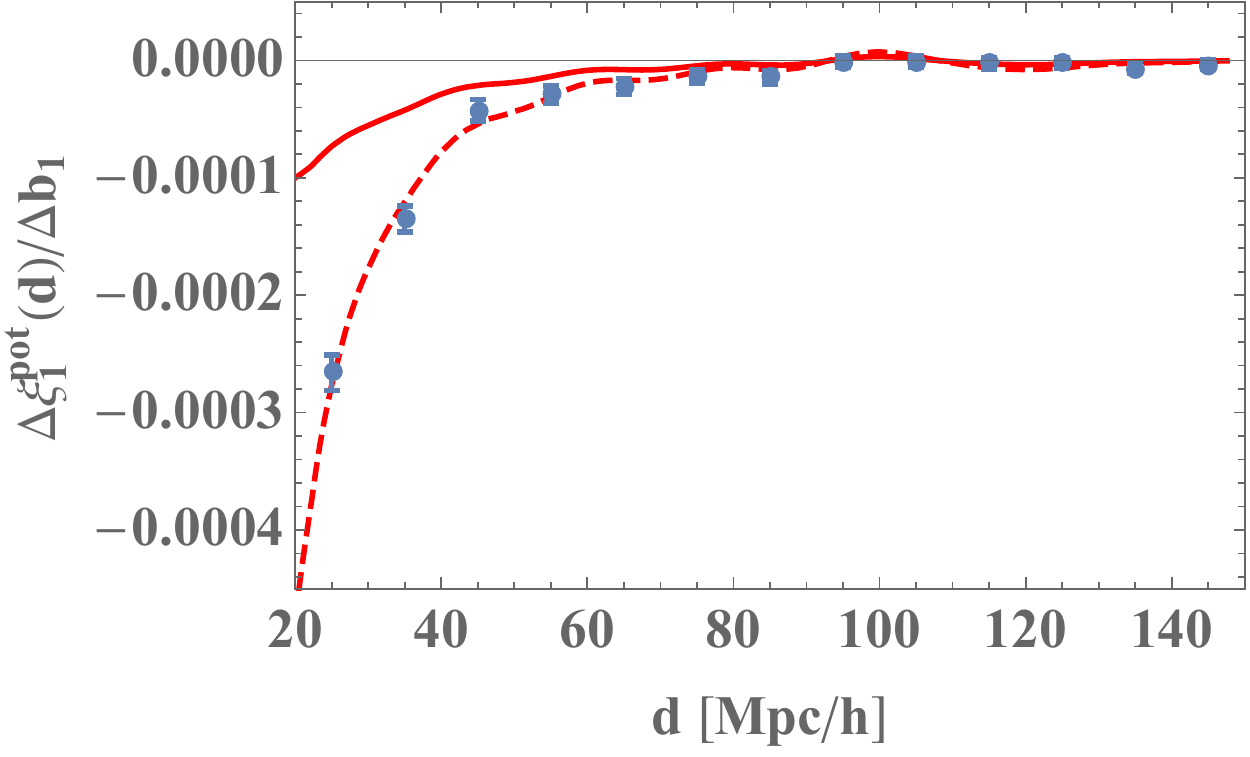} 
\caption{ We plot the full next-to-leading order dipole of the correlation function (left panel) and the contribution induced by the gravitational potential (right panel) at $z=0.341$. We compare it with the numerical results of Ref.~\cite{Breton:2018wzk} (from which we have subtracted their linear theoretical predictions).
Dashed (solid) lines include (neglect) the non-linear velocity dispersion parameter, see Appendix~\ref{app:veldisp}.
}
\label{fig_full_dip}
\end{center}
\end{figure}
 \begin{figure}[h!]
\begin{center}
\includegraphics[width=0.45\textwidth]{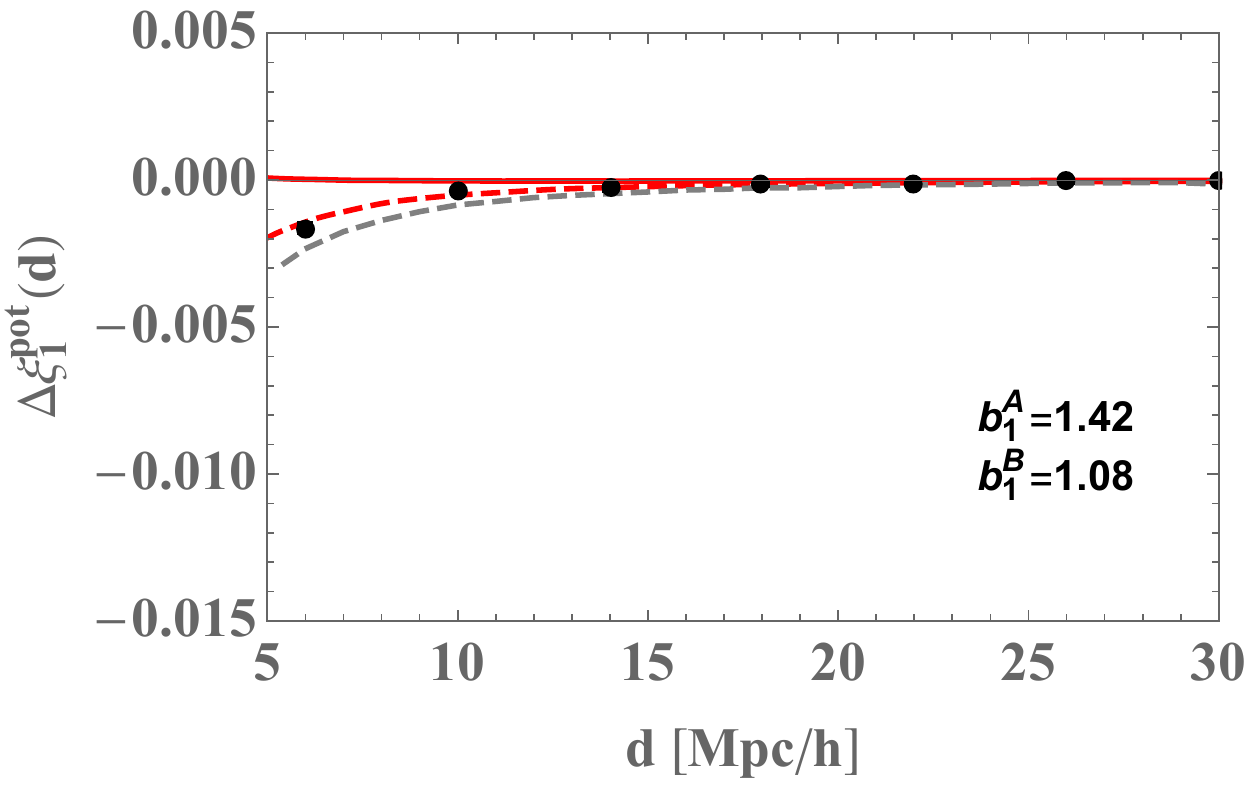} 
\includegraphics[width=0.45\textwidth]{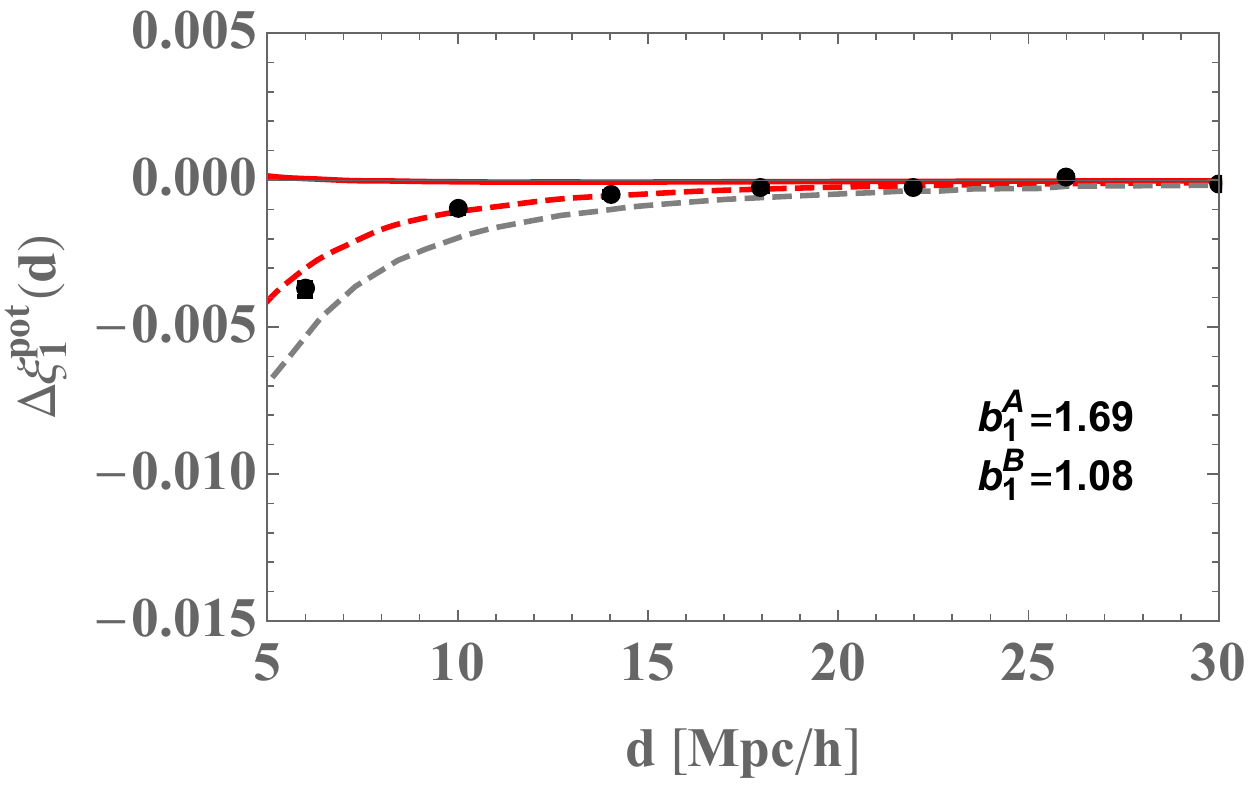} \\
\includegraphics[width=0.45\textwidth]{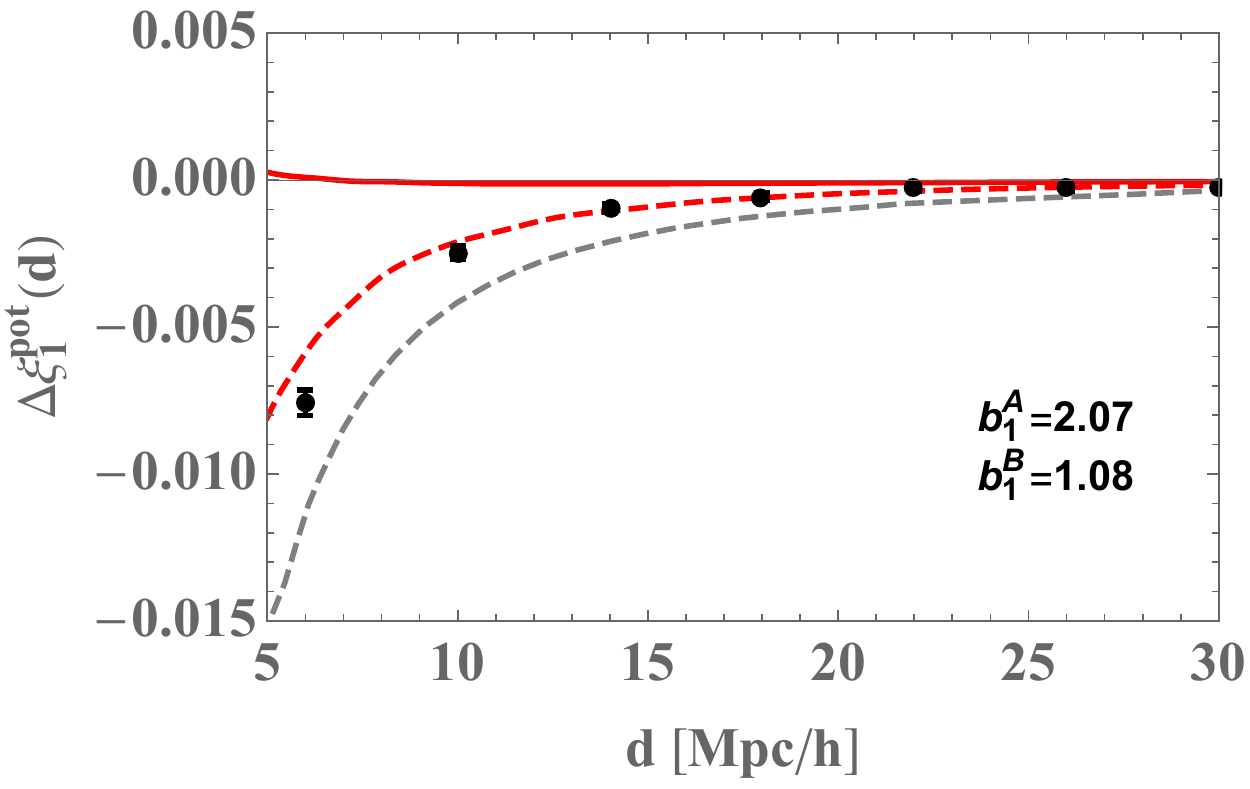} 
\includegraphics[width=0.45\textwidth]{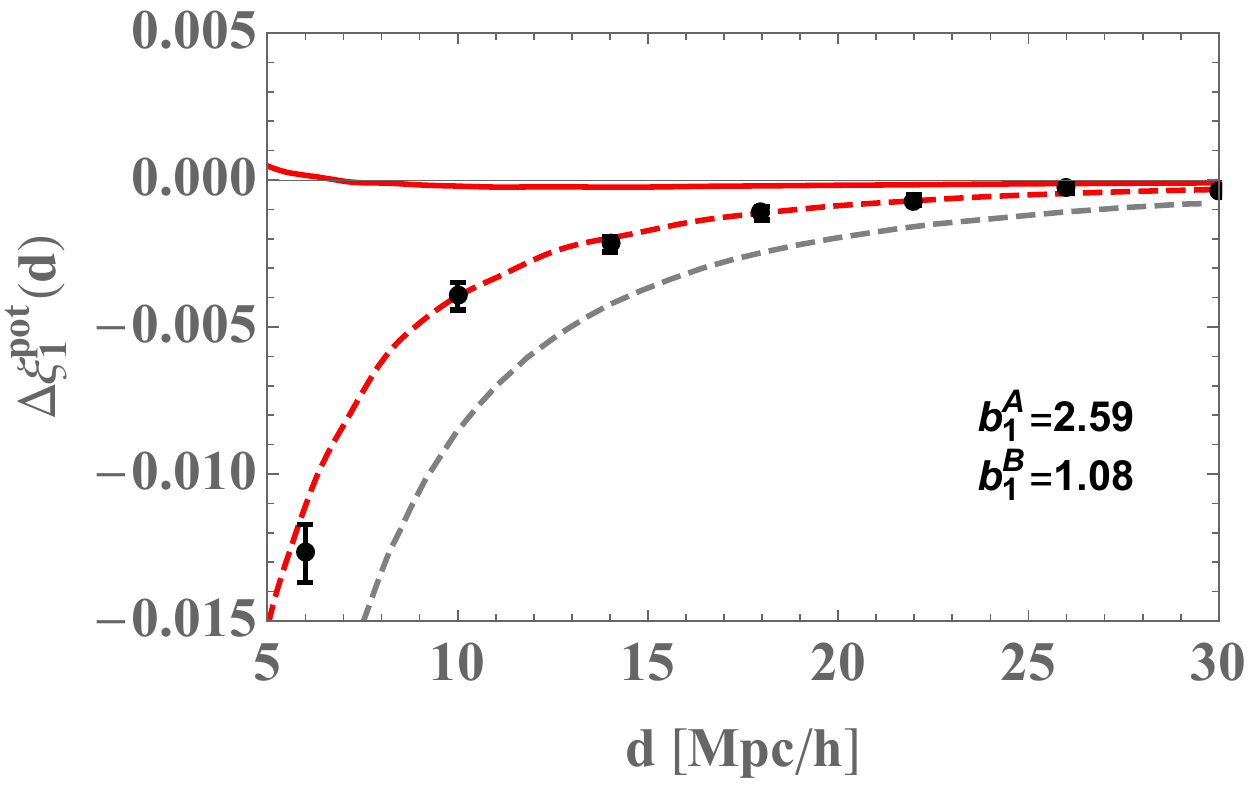} 
\caption{ We plot the contribution of the gravitational terms to the dipole of the correlation function at 1-loop. The different panels represent different bias combinations at $z=0.341$. We compare our results (red lines) with Ref.~\cite{Breton:2018wzk}. 
Red dashed (solid) lines include (neglect) the non-linear velocity dispersion parameter, see Appendix~\ref{app:veldisp}.
The gray dashed line is the theoretical prediction of gravitational redshift following Ref.~\cite{Croft:2013taa}, as computed in Ref.~\cite{Breton:2018wzk}. In the comparison with the numerical results of Ref.~\cite{Breton:2018wzk} we have subtracted their linear theoretical predictions.}
\label{fig_diffbias}
\end{center}
\end{figure}
In fig.~\ref{fig_full_dip} we compare with Ref.~\cite{Breton:2018wzk} the full next-to-leading order dipole (left panel) and the contribution induced by the gravitational potential (right panel). We remark that we neglect some effects with respect to the numerical analysis of Ref.~\cite{Breton:2018wzk}, as for instance the evolution effects. This may impact the accuracy of the full dipole comparison. If we focus only on the gravitational potential term, our results agree with very good accuracy with the results based on light-tracing N-body simulation developed in Ref.~\cite{Breton:2018wzk}, by introducing a single EFT-inspired parameter in order to capture the non-perturbative behaviour on small scales (see Appendix~\ref{app:veldisp}).
 To confirm that our formalism is capturing also the correct bias dependence we further compare, see fig.~\ref{fig_diffbias}, the gravitational redshift effect for different combinations of bias values. Indeed, while on linear scale the bias expansion is described by the linear bias $b_1$, on smaller scales a proper description requires a higher order bias expansion. Due to the relation between the different bias parameters, provided by the numerical fit~\eqref{bias_relation}, the dipole of the correlation function is not simply proportional to the bias difference beyond linear scales, but it depends explicitly on the biases of the two galaxy populations.

The good agreement with simulations provides a strong support in favour of the correctness of the 1-loop corrections we have computed in our work.
In particular, our results seem in a better agreement with numerical predictions with respect to previous analytical model of Ref.~\cite{Croft:2013taa}, which seems to
overestimate the amplitude by about a factor of two, as shown in Ref.~\cite{Breton:2018wzk}.
 It is worth remarking that in our work, and in Ref.~\cite{Breton:2018wzk}, we include all the terms induced by the gravitational potential on the dipole of the correlation function. These terms are derived in a relativistic framework by solving the geodesic equations.
 In Ref.~\cite{Croft:2013taa} and afterwards in Refs.~\cite{Alam:2017cja,Alam:2017izi,Zhu:2017jfl} the Authors simply consider the difference in the gravitational potential at different source positions, by ignoring all the other relativistic contributions involving the gravitational potential.

It is important to remark that in order to achieve a good agreement between our perturbative approach and numerical simulations, we need to include the velocity dispersion as predicted by the non-linear matter power spectrum. The correction to the velocity dispersion can be interpreted as an EFT parameter and its value fitted with simulations, see Appendix~\ref{app:veldisp} for detailed implementation. In our work we show that with a single free parameter (to account for the non-perturbative effects on small scales) we achieve a good agreement on a large range of scales.

\section{Conclusions}
\label{sec:conclusions}
\vspace{0pt}
In this work we derived the dipole of the correlation function at 1-loop including the relevant relativistic effects. We computed for the first time the leading relativistic corrections to third order in perturbation theory. Our results agree with very good accuracy with previous numerical results based on N-body simulations~\cite{Breton:2018wzk}. This provides a solid confirmation of the validity of our perturbative approach. To achieve this agreement we need to introduce an EFT-inspired free parameter, that we fit with simulations provided by Ref.~\cite{Breton:2018wzk} in order to capture the non-perturbative effects on small scales.
Since the relativistic effects are induced by the perturbative solution of geodesic equations, the range of validity of our perturbative expansion is limited to the weak field approximation\footnote{We also assume the local bias expansion.}.

Our results show that if we assume the Euler equation dictated by the Equivalence Principle, the amplitude of the dipole sourced by the gravitational potential is comparable to, or smaller, than other effects 
induced by peculiar velocities, at least on large scales where the PT approach is valid (scales larger than $5-10{\rm Mpc/h}$). This complicates the interpretation of these measurements, as the measurement 
of gravitational redshift from LSS~\cite{Alam:2017izi}. Indeed, the amplitude of several  peculiar velocity contributions to the dipole (at any order) depends on different bias coefficients (clustering, magnification and evolution) of the two populations. Therefore, an unbiased measurement of the gravitational redshift effect on LSS requires an accurate prior knowledge of all these bias parameters. For this purpose, the measurement of the linear dipole can provide some useful information on magnification and evolution biases.

We stress that due to the cancellation between the gradient of the gravitational potential and the acceleration of galaxy motion induced by the Euler equation, the amplitude of the 2-point function proportional to the linear bias is suppressed and therefore a non-linear bias model is required to properly describe the gravitational redshift effect.

\section*{Acknowledgements}
It is a pleasure to thank Michel-Andrès Breton for clarifying and sharing their numerical results. We thank Elena Giusarma for useful discussions.
ED is supported by the Swiss National Science Foundation (No.~171494).

\appendix
\section{Derivation of second order number counts}
\label{app:second}
We derive the relativistic number counts to second order in perturbation theory. Differently from previous derivations we consider terms only up to first order in the relativistic expansion parameter $\HH/k$ neglecting integrated terms. This at the same time simplifies the cumbersome calculation to second order (see previous results in literature~\cite{Yoo:2014sfa,DiDio:2014lka,Bertacca:2014dra}) and captures the relevant terms which lead to a non-vanishing dipole.

To derive the galaxy number counts in terms of observable quantities $\bn$ and $z$, we need to perform a change of coordinates. Considering that the density of sources transforms as a scalar density field we have to combine a density and a volume part transformations. To linear order this is simply given by a linear combination of them
\be
\Delta^{(1)} \left( \bn , z \right) = \delta_z \left( \bn , z \right) + \frac{\delta V^{(1)} \left( \bn , z \right) }{\bar V \left( z \right)}
\ee
where $ \delta_z \left( \bn , z \right)  $ denotes the density redshift and $\delta V^{(1)} \left( \bn , z\right)$ the volume perturbations, whereas $\bar V\left( z \right)$ is the background volume element.
At second order (omitting the subscript~$^{(1)}$ when obvious) we have
\bea
\Delta^{(2)} \left( \bn, z \right) &=&  \delta^{(2)}_z \left( \bn , z \right) +  \frac{\delta V^{(2)} \left( \bn , z \right)}{\bar V \left( z \right)} + \delta_z \left( \bn , z \right)  \frac{\delta V \left( \bn , z \right)}{\bar V \left( z \right)}
\nonumber \\
&&
- \langle \delta^{(2)}_z \left( \bn , z \right) \rangle - \langle  \frac{\delta V^{(2)} \left( \bn , z \right)}{\bar V \left( z \right)} \rangle - \langle \delta_z \left( \bn , z \right)  \frac{\delta V \left( \bn , z \right)}{\bar V \left( z \right)}  \rangle \, .
\label{counts2}
\eea
We consider a metric in Poisson gauge
\be \label{metric}
ds^2 = a^2 \left[ - \left( 1 + 2 \Psi \right) dt^2  + \left( 1 -2 \Phi \right) \left(  dr^2 + r^2 d\Omega \right) \right]
\ee
where we expand the metric perturbations as
\be
\Psi= \Psi^{(1)} + \Psi^{(2)} + \Psi^{(3)} \qquad \text{and} \qquad \Phi= \Phi^{(1)} + \Phi^{(2)} + \Phi^{(3)} \, .
\ee
Different gauges will only induce terms suppressed at least by $\left( \HH/k \right)^2$ and therefore negligible in our approximation.

We remark that we do not consider perturbations of the line of sight direction $\bn$. It is well known that perturbations of the line of sight are induced by the gradient of the lensing potential and they affect the matter power spectrum, see Ref.~\cite{Dodelson:2008qc,DiDio:2016kyh}. Nevertheless this is orthogonal to the line of sight direction and therefore we do not expect to produce any relevant asymmetry along the radial direction which can source a dipole. This is also consistent with our approximation scheme which neglects metric perturbation integrated along the line of sight.
Relativistic corrections to galaxy clustering involving gravitational potential terms have been computed beyond linear theory in Refs.~\cite{Yoo:2014sfa,Bertacca:2014dra,DiDio:2014lka} (at second order) and~\cite{Nielsen:2016ldx,Jalivand:2018vfz} (at third order).

\subsection{Density perturbation}
We start considering the density redshift perturbation by comparing the observed fluctuation of number of galaxy with its theoretical description
\be
\delta_z \left( \bn , z \right) \equiv \frac{\rho_\text{obs} \left( \bn , z \right) - \langle \rho_\text{obs} \left( \bn,z \right) \rangle }{\langle \rho_\text{obs} \left(\bn, z \right) \rangle}
=
\frac{ \rho \left(\bn, \bar z \right)  - \langle\rho \left(\bn, \bar z \right) \rangle}{\langle \rho \left( \bn,\bar z \right) \rangle}
\ee
where $\bar z$ represents a background redshift $1+ \bar z= 1/a(t)$ of a theoretical model built to fit the observations. To second order we obtain
\bea
\delta^{(1)}_z \left( \bn, z \right) +\delta_z^{(2)} \left( \bn, z \right) &=&\frac{\bar \rho \left( z - \delta z^{(1)} - \delta z^{(2)} \right) + \delta\rho^{(1)} \left( \bn , z - \delta z^{(1)} \right) + \delta\rho^{(2)} \left( \bn , z \right) }{\langle \bar \rho \left( z - \delta z^{(1)} - \delta z^{(2)} \right) + \delta\rho^{(1)} \left( \bn , z - \delta z^{(1)} \right) + \delta\rho^{(2)} \left( \bn , z \right)\rangle } -1
\nonumber \\
&& \hspace{-4.5cm}=
\frac{\bar \rho \left(  z \right) \!-\! \frac{d\bar \rho}{d z} \left( \delta z^{(1)} \!+ \!\delta z^{(2)} \right) \!+ \!\frac{1}{2} \frac{d^2 \bar \rho}{dz^2} \left( \delta z^{(1)} \right)^2 \!+ \!\delta \rho^{(1)} \left( \bn, z \right)\!-\! \frac{d \delta\rho^{(1)}}{dz} \delta z^{(1)}\! +\! \delta \rho^{(2)} \left( \bn , z \right) }{
\bar \rho \left( z \right)
 + \frac{1}{2} \frac{d^2 \bar \rho}{dz^2} \langle \left( \delta z^{(1)} \right)^2 \rangle -  \frac{d\bar \rho}{dz} \langle \delta  z^{(2)} \rangle  - \langle \frac{d \delta\rho^{(1)}}{dz} \delta z^{(1)} \rangle + \langle \delta \rho^{(2)} \left( \bn , z \right)  \rangle} -1 \, ,
 \quad
 \label{eq.A6}
\eea
and therefore
\bea
\label{delta1}
\delta_z^{(1)} \left( \bn , z \right) &=& \delta^{(1)} - \frac{1}{\bar\rho} \frac{d \bar\rho}{d\bar z} \delta z^{(1)} \, , \\
\delta_z^{(2)} \left( \bn , z \right) &=& \delta^{(2)} - \frac{d \delta}{d \bar z} \delta z - \frac{1}{\bar \rho} \frac{d\bar \rho}{dz} \delta \ \delta z + \frac{1}{2\bar \rho} \frac{d^2 \bar \rho}{dz^2} \left( \delta z \right)^2 - \frac{1}{\bar \rho} \frac{d \bar \rho}{dz} \delta z^{(2)}
\nonumber \\
&&
- \langle \delta^{(2)} - \frac{d \delta}{d \bar z} \delta z - \frac{1}{\bar \rho} \frac{d\bar \rho}{dz} \delta^{(1)} \delta z + \frac{1}{2\bar \rho} \frac{d^2 \bar \rho}{dz^2} \left( \delta z \right)^2 - \frac{1}{\bar \rho} \frac{d \bar \rho}{dz} \delta z^{(2)}  \rangle \, .
\label{delta2}
\eea
We need at this point to compute the redshift perturbation, starting from the redshift definition
\be
1 + z = \frac{\left. k^\mu u_\mu \right|_s}{\left. k^\mu u_\mu \right|_o}
\ee
where the 4-velocity field is determined by
\be
\left( u^\mu \right) = a^{-1} \left( 1- \psi^{(1)}+ \frac{1}{2}  { v}^2 + \frac{3}{2} \psi^2 - \psi^{(2)} , {\bf v}^{(1)} +{\bf v}^{(2)} \right) \, ,
\ee
such that $u^\mu u_\mu = -1$, and $k^\mu= d x^\mu/d\lambda$ is the light-like 4-vector along the photon geodesic path and $\lambda$ is the associated affine parameter. Before solving the geodesic equation to obtain $k^\mu$, we derive also the volume contribution in terms of redshift perturbation $\delta z$. This helps us to understand at which order in the relativistic expansion parameter $k/\HH$ we need to solve the geodesic equations.

\subsection{Volume perturbation}
Following the approach of Ref.~\cite{Bonvin:2011bg} we start considering  an infinitesimal volume element around the source defined by\footnote{The angles $\theta$ and $\varphi$ describe the angular position in the sky of the source: $\bn= \bn \left( \theta, \varphi \right)$}
\be
dV = \sqrt{-g} \epsilon_{\mu \nu \alpha \beta} u^\mu dx^\nu dx^\alpha dx^\beta = v\left(z, \theta, \varphi \right) dz d\theta d\varphi
\ee
where
\bea
\label{derivation_volume}
v_\text{obs}\left(z, \theta, \varphi \right) &=& \sqrt{-g} \epsilon_{\mu \nu \alpha \beta} u^\mu \frac{ \partial x^\nu}{\partial z}\frac{ \partial x^\alpha}{\partial \theta} \frac{ \partial x^\beta}{\partial \varphi}
\nonumber \\
&=&
\left[a^3 r^2 \sin\theta\right](\bar z) \left\{\left( 1 + \frac{\ndv^2}{2}+  \frac{v_\perp^2}{2} \right) \frac{dr(\bar z)}{dz} + \left( \ndv^{(1)} (\bar z)+ \ndv^{(2)} (\bar z)\right) \frac{dt(\bar z)}{dz} \right\}
\nonumber \\
&=&
\left[ a^3 r^2 \sin\theta \right] (\bar z) \left\{ \frac{dr}{d \bar z} \left( 1 - \frac{d\delta z^{(1)}}{d z} - \frac{d\delta z^{(2)}}{d z}+ \frac{\ndv^2}{2} +  \frac{v_\perp^2}{2}  \right)
\right.
\nonumber \\
&& \qquad\qquad\qquad
\left.
+ \left( \ndv^{(1)}- \frac{d\ndv}{dz}  \delta z+ \ndv^{(2)}  \right) \frac{dt}{d\bar z} \left( 1 - \frac{d \delta z^{(1)}}{dz} \right) \right\}
\nonumber \\
&=&
\left[ \frac{a^4 r^2 \sin\theta}{\HH} \right] (\bar z) \left\{\left( 1 - \frac{d\delta z^{(1)}}{d z} - \frac{d\delta z^{(2)}}{d z }   + \frac{\ndv^2}{2}  +  \frac{v_\perp^2}{2}  \right)
\right.
\nonumber \\
&& \qquad\qquad\qquad
\left.
- \left( \ndv^{(1)}- \frac{d \ndv}{dz}  \delta z^{(1)} + \ndv^{(2)}  \right) \left( 1 - \frac{d \delta z^{(1)}}{dz} \right) \right\}
\nonumber \\
&=&
\bar v (z) \left( 1 -\frac{1}{\bar v}\frac{dv}{dz} \left( \delta z^{(1)} + \delta z^{(2)} \right) + \frac{1}{2 \bar v} \frac{d^2 \bar v}{dz^2} \left( \delta z \right)^2 \right)
\nonumber \\
&&
\left( 1- \frac{d\delta z^{(1)}}{dz} - \frac{d \delta z^{(2)}}{dz}   + \frac{\ndv^2}{2} + \frac{v_\perp^2}{2}  - \ndv^{(1)} +  \ndv \frac{d\delta z}{dz} + \frac{d\ndv}{dz}  \delta z^{(1)} - \ndv^{(2)} \right) \, ,
\nonumber \\
\label{vol2}
\eea
where we have introduced $\bar v =a^4 r^2  \HH^{-1} \sin\theta$.
In this derivation we have already neglected the subleading contribution of the metric perturbation in the determinant of the metric $g$ and in the 4-velocity $u^\mu$. Linear order metric perturbations $\Psi$ and $\Phi$ are suppressed with respect to $\delta^{(2)}$ by $\left( \HH/k \right)^4$ and second order gravitational potential $\Psi^{(2)}$ and $\Phi^{(2)}$ by $\left( \HH/k \right)^2$. We also did not consider the angular Jacobian 
\be
\left| \frac{\partial \left( \theta_s, \varphi_s \right)}{\partial \left( \theta_o , \varphi_o \right)} \right|,
\ee
since this term leads to the lensing magnification and we neglect it in the current derivation.

Considering that we can rewrite the derivative with respect to the redshift as
\be
\frac{d}{dz} = \frac{dt  }{dz} \partial_t + \frac{dx^i}{dz} \partial_i 
\ee
we observe that we need to derive $\delta z$ at least at the order $\left( \HH/ k \right)^2$. Clearly for the terms which involve an integral along the line of sight we need to consider only the ones at least of the order $\HH/k$.

\subsection{Redshift perturbation}
In order to find the redshift perturbation we need to solve the geodesic equation 
\be
\frac{d k^\mu}{d \lambda} + \Gamma^{\mu}_{\nu \rho} k^\nu k ^\rho =0 \, ,
\ee
where $ \Gamma^{\mu}_{\nu \rho}$ are the Christoffel symbol of the metric~\eqref{metric}. At the background we simply have
\be
\left( \bar k^\mu \right) = \frac{1}{a(t)^2} \left( 1, -1, 0 , 0 \right) \, .
\ee
To linear order we first remark that $k_\perp^a$ is described by the transversal gradient of the lensing potential. As previously motivated, we do not consider perturbation of the line of sight direction $\bn$ and therefore we set $k_\perp^a=0$ to any order. As we will see at the end of this section, this ansatz will reproduce the leading terms of the linear~\cite{,Bonvin:2011bg,Challinor:2011bk} and second order~\cite{DiDio:2014lka} galaxy number counts derived in the literature.
Hence, we find to first order 
\be
\left(  k^\mu \right)^{(1)} \simeq \frac{1}{a(t)^2} \left( -2 \Psi,\Psi-\Phi , 0 , 0 \right)
\ee
and to second order
\be
\left(  k^\mu \right)^{(2)} \simeq \frac{1}{a(t)^2} \left( -2 \Psi^{(2)},\Psi^{(2)}-\Phi^{(2)} , 0 , 0 \right)\, .
\ee

By normalizing the affine parameter $\lambda$ such that $\left. u^\mu k_\mu\right|_o =-1$ we compute the redshift perturbation
\bea
1+z &=& - \left.   k^\mu u_\mu \right|_s \simeq \left(1+ \bar z \right) \left( 1+ \left(  k^\mu u_\mu\right)^{(1)} \left( \bar z \right) + \left( k^\mu u_\mu\right)^{(2)} \left( \bar z \right)
\right)
\nonumber 
\\
&\simeq& \left( 1 + z - \delta z ^{(1)} - \delta z^{(2)} \right) \left( 1+ \left(  k^\mu u_\mu \right)^{(1)} \left(  z \right) - \delta z^{(1)} \frac{d}{dz}\left(  k^\mu u_\mu\right)^{(1)} \left(  z \right)  +\left(  k^\mu u_\mu \right)^{(2)} \left(  z \right)
\right) \, ,
\nonumber \\
\eea
from which it follows
\bea
\delta z^{(1)} &=& \left( 1+z \right)  \left(  k^\mu u_\mu \right)^{(1)} \left(  z \right) 
\simeq - \left( 1+z \right) \left( \Psi + \ndv \right) \, ,
\\
\delta z^{(2)} &=& \left( 1+z \right)\left(  \left(  k^\mu u_\mu \right)^{(2)} \left(  z \right) - \delta z^{(1)} \frac{d}{dz}\left(  k^\mu u_\mu\right)^{(1)} \left(  z \right)  \right) 
\nonumber  \\
&\simeq&
- \left( 1+z \right) \left( -\frac{1}{2} \ndv^2 +{\Psi^{(2)}}- \frac{1}{2} v_{\perp}^2 + \ndv^{(2)}  - \frac{d\ndv}{d  z }  \delta z {-\frac{d\Psi}{d  z }  \delta z} \right) + \ndv \delta z + {\Psi \delta z}
\nonumber\\
&\simeq& - \left( 1+z \right) 
\left( \ndv^{(2)}+ \Psi^{(2)} - \frac{v^2}{2} +  \HH^{-1} \ndv\partial_r \Psi  + \ndv^2 \right) \, .
\eea
Combining all the parts together we find Eq.\eqref{Delta2N} and 
\bea
\label{Delta2R_noE}
\Delta_R^{(2)} &=&  \left( -1 +\frac{\dot \HH}{\HH^2} + \frac{2}{\HH r} - f_{\rm evo}\right) \ndv^{(2)} 
+ \left( -2 +3 \frac{\dot \HH}{\HH^2} + \frac{4}{\HH r} - 2 f_{\rm evo} \right) \HH^{-1} \ndv \partial_r \ndv
\nonumber \\
&&
+ \left( -1+ \frac{\dot \HH}{\HH^2} + \frac{2}{\HH r} - f_{\rm evo} \right) \ndv \delta 
- \HH^{-1} \dot\ndv^{(2)} - 2 \HH^{-2} \partial_r \ndv \dot \ndv - 2 \HH^{-2} \ndv \partial_r \dot \ndv
\nonumber \\
&&
 - \HH^{-1} \dot \ndv \delta - \HH^{-1} \ndv \dot \delta +   \HH^{-1}  v^a \partial_a \ndv  
 + \HH^{-2}\Psi\partial_r^2 \ndv  + \HH^{-1} \Psi \partial_r \delta
\nonumber \\
&&
{
 + \HH^{-1} \partial_r \Psi^{(2)}  + 2 \HH^{-2} \partial_r \ndv \partial_r \Psi+ \HH^{-1} \delta \partial_r \Psi + \HH^{-2} \ndv \partial_r^2 \Psi
} \,.
\eea
We then assume that galaxies move along geodesics as a pressure-less fluid satisfying the Euler equation\footnote{Consistently with the approximation scheme used in our work we consider only terms up to $\HH/k$.}
\bea 
\label{Euler_eq_1}
\dot \ndv^{(1)} + \HH \ndv^{(1)}  { - \partial_r \Psi}&=& 0 \, , \\
\dot \ndv^{(2)} + \HH \ndv^{(2)} - \ndv \partial_r \ndv + v^a_\perp \partial_a \ndv  { - \partial_r \Psi^{(2)}} &\simeq& 0 \, .
\label{Euler_eq_2}
\eea
Combining these with Eq.~\eqref{Delta2R_noE} and adding magnification bias as shown in Appendix~\ref{sec:s_bias}, we obtain Eq.~\eqref{Delta2R}.

\section{Derivation of third order number counts}
\label{app:third}
In this section we derive the number counts to third order in perturbation theory including all the relativistic effects up to linear order in $\HH/k$. We follow and extend the same procedure successfully used to second order in the previous section. We can write the number counts as
\be
\Delta^{(3)} \left( \bn , z \right)  = \delta_z^{(3)} \left(\bn , z \right) + \delta_z^{(2)} \left( \bn,z \right) \frac{\delta V\left( \bn , z\right) }{\bar V \left( z \right) } + \delta_z \left( \bn,z \right)\frac{\delta V^{(2)}\left( \bn , z\right) }{\bar V \left( z \right) }  +\frac{\delta V^{(3)}\left( \bn , z\right) }{\bar V \left( z \right) } \, .
\ee
By extending Eq.~\eqref{eq.A6} up to third order we find 
\bea
\delta_z^{(3)} \left( \bn , z \right) &=&
\delta^{(3)} - \left( \delta^{(2)} \delta z^{(1)} + \delta^{(1)} \delta z^{(2)} + \delta z^{(3)} - \delta z^2 \frac{d \delta}{dz}  \right)\frac{1}{\bar \rho} \frac{d \bar \rho}{ dz}  - \delta z^{(2)} \frac{d \delta}{dz}
\nonumber \\
&&+ \left( \frac{1}{2} \delta^{(1)} \delta z^2 + \delta z \delta z^{(2)} \right)\frac{1}{\bar \rho} \frac{d^2 \bar \rho}{ dz^2}  - \frac{1}{6} \delta z^3 \frac{1}{\bar \rho} \frac{d^3 \bar \rho}{ dz^3} 
- \delta z \frac{d \delta^{(2)}}{dz} + \frac{1}{2} \delta z^2 \frac{d^2 \delta}{dz^2} \, .
\quad
\eea
In order to compute the volume and the redshift perturbations we need to extend the peculiar velocity to third order, under the normalization condition $u^\mu u _\mu =-1$, 
\be
\left( u^\mu \right)^{(3)} = a^{-1} \left( -\Psi^{(3)} - v^2 \left( \Phi+ \frac{\Psi}{2} \right) +3 \Psi \Psi^{(2)} -\frac{5}{2} \Psi^3+ \bv \cdot \bv^{(2)} , {\bf v}^{(3)} \right) \, .
\ee
Following Eq.~\eqref{vol2} we find the volume perturbation
\bea
\frac{\delta V^{(3)} \left( \bn , z \right) }{\bar V \left( z \right) } & \simeq &
- \frac{d \delta z^{(3)}}{dz}
 - \ndv^{(3)} + \left( 1+ z \right)^{-1} \left( 4 - \frac{\dot \HH}{\HH^2}- \frac{2}{\HH r} \right) \left( \delta z^{(3)} - \delta z^{(2)} \frac{d \delta z}{dz} - \delta z \frac{d \delta z^{(2)}}{dz} 
 \right)
 \nonumber \\
&&
 \left( 1+ z \right)^{-1} \HH^{-1} \left( 
 \partial_r \ndv^{(2)} \delta z + \partial_r \ndv \delta z^{(2)} - \partial_r \ndv  \delta z \frac{d \delta z}{dz}
 \right)
 \nonumber \\
&&
 + \ndv^{(2)} \frac{d \delta z }{dz} + \ndv \frac{d \delta z^{(2)} }{dz}
 - \frac{\delta z^2}{2 \HH^2 \left( 1+ z\right)^2} \partial_r^2 \ndv \, .
\eea
Then, by solving the geodesic equation to third order we find at the relevant order in relativistic correction
\be
\left(  k^\mu \right)^{(3)} \simeq \frac{1}{a(t)^2} \left( -2 \Psi^{(3)},\Psi^{(3)}-\Phi^{(3)} , 0 , 0 \right)\, .
\ee
Observing that only the time component of $\left(  k^\mu \right)^{(3)}$ plays a role in determining the following redshift perturbation, we obtain
\bea
\delta z^{(3)} &=& \left( 1+z \right)\left(  \left(  k^\mu u_\mu \right)^{(3)} \left(  z \right) - \delta z^{(2)} \frac{d}{dz}\left(  k^\mu u_\mu\right)^{(1)} \left(  z \right) - \delta z^{(1)} \frac{d}{dz}\left(  k^\mu u_\mu\right)^{(2)} \left(  z \right)  
\right.
\nonumber \\
&& \qquad \qquad
\left. 
+\frac{1}{2} \delta z^2 \frac{d^2 }{dz^2} \left(  k^\mu u_\mu\right)^{(1)} 
\right) 
\nonumber \\
&\simeq&
- \left( 1+ z \right) \left( \ndv^{(3)}  + \Psi^{(3)} - \bv^{(1)} \cdot \bv^{(2)} + \HH^{-1} \ndv \partial_r \Psi^{(2)} + \frac{1}{2 \HH^2} \ndv^2 \partial^2_r \Psi 
\right. 
\nonumber \\
&&
\left. \qquad \qquad \quad
+ \HH^{-1} \ndv^{(2)} \partial_r \Psi
 + 2 \ndv \ndv^{(2)} \right) \, .
\eea

Combining these results we find Eq.~\eqref{Delta3N} and 
\bea
\Delta_R^{(3)} &=&
\frac{1}{2 \HH^3} \partial^3_r \left( \ndv^2 \Psi \right)
 - \frac{1}{2 \HH^3} \partial_t \partial_r^2 \ndv^3
 + \frac{1}{\HH^2} \partial_r^2 \left( \Psi \ndv \delta \right)
 + \frac{1}{\HH^2}\left[ \partial_r^2 \left( \Psi \ndv  \right) \right]^{(3)}
 - \frac{1}{2\HH^2} \partial_r^2 \left( \ndv v^2 \right)
 \nonumber \\
 &&
 - \frac{2}{\HH^2} \partial_t \partial_r \left( \ndv \ndv^{(2)} \right)
 - \frac{1}{\HH^2} \partial_t \partial_r \left( \delta \ndv^2 \right)
 - \frac{1}{\HH} \partial_r \left( \bv \cdot \bv^{(2)} \right)
 - \frac{1}{2 \HH} \partial_r \left( \delta v^2 \right)
 +\frac{1}{\HH} \left[ \partial_r \left( \delta \Psi \right) \right]^{(3)}
  \nonumber \\
 &&
 + \frac{1}{\HH} \partial_r \Psi^{(3)}
 - \frac{1}{\HH} \left[ \partial_t \left( \delta \ndv \right) \right]^{(3)}
 - \frac{1}{\HH} \dot \ndv^{(3)}
 + \left( \ndv^{(3)} + \left[ \delta \ndv \right]^{(3)} \right) \left( - 1 + \frac{\dot \HH}{\HH^2} + \frac{2}{\HH r } - f_{\rm evo} \right)
   \nonumber \\
 &&
 + \frac{1}{\HH^2} \partial_r \left( \partial_r \ndv \ndv^2 \right) \left( 3 \frac{\dot\HH}{\HH^2} + \frac{3}{\HH r} - \frac{3}{2} f_{\rm evo} \right) 
 + \frac{1}{\HH} \partial_r \left( \ndv \ndv^{(2)} \right) \left( - 1 + 3\frac{\dot\HH}{\HH^2} + \frac{4}{\HH r} - 2 f_{\rm evo} \right)
    \nonumber \\
 &&
 \frac{1}{\HH} \partial_r \left( \delta \ndv^2 \right) \left( - \frac{1}{2} + \frac{3}{2} \frac{\dot\HH}{\HH^2} + \frac{2}{\HH r} - f_{\rm evo} \right) \, .
\eea
Then by using Euler Eqs.~\eqref{Euler_eq_1},~\eqref{Euler_eq_2} and 
\be
\label{Euler_eq_3}
\dot \ndv^{(3)} + \HH \ndvthree + \left( v^{(2)} \right)^a \partial_a \ndv  + v^a \partial_a \ndvtwo - \ndv \partial_r \ndvtwo - \ndvtwo \partial_r \ndv - \partial_r \Psi^{(3)} \simeq 0 
\ee
we obtain Eq.~\eqref{Delta3R}.

\section{Magnification bias}
\label{sec:s_bias}
The magnification bias accounts for the fact that we observe only galaxies above a given luminosity threshold $L$, i.e
\be \label{eq:c1}
n \left( \bn, z , >L \right) = n \left( \bn, z , >\bar L \right) + \Delta L \left[ \frac{\partial n}{ \partial L}  \right]_{L = \bar L} + \frac{1}{2}  \Delta L^{2} \left[ \frac{\partial^2 n}{ \partial L^2}  \right]_{L = \bar L}
 + \frac{1}{6}  \Delta L^{3} \left[ \frac{\partial^3 n}{ \partial L^3}  \right]_{L = \bar L} \, .
\ee
At constant flux the luminosity and the distance are related through
\be
L \propto d_A^2 \, ,
\ee
therefore ($\delta L^{(i)} = \Delta L^{(i)} / \bar L$ and $\delta d^{(i)} = \Delta d^{(i)} / \bar d$ )
\bea
\delta L^{(1)} &\rightarrow& 2 \delta d^{(1)} , \\
\delta L^{(2)} &\rightarrow& \left( \delta d^{(1)}\right)^2 +2 \delta d^{(2)}, \\
\delta L^{(3)} &\rightarrow& 2 \delta d^{(1)} \delta d^{(2)} +2 \delta d^{(3)} \, .
\eea
We consider that the most relevant contribution to the distance is given by the peculiar motion of the source
\be \label{dist_vel}
\delta d^{(i)} \simeq \left( \frac{1}{\HH r} - 1 \right) \ndv^{(i)} \, .
\ee
Hence only the first order in the Taylor expansion~\eqref{eq:c1} leads to relevant terms.
This leads to
\bea
\Delta^{(1)} \left( \bn , z , >L \right) &=&
 \Delta^{(1)} \left( \bn , z , > \bar L \right) + 2  \left( 1 - \frac{1}{\HH r} \right)Q \ndv \, ,
\\
\Delta^{(2)} \left( \bn , z , >L \right) &\simeq& 
\Delta^{(2)} \left( \bn , z , > \bar L \right) 
+2 \delta d^{(1)} \left( \frac{\partial \Delta^{(1)}}{\partial \ln L} \right)_{L = \bar L}
-2 Q \left( \delta d^{(2)} +  \delta d^{(1)} \Delta^{(1)}  \right)
\nonumber \\
&\simeq&
\Delta^{(2)} \left( \bn , z , > \bar L \right) 
+ 2  \left( 1 - \frac{1}{\HH r} \right)Q 
\left(
\ndvtwo + \delta_g \ndv + \frac{\ndv}{\HH} \partial_r \ndv 
 \right) 
 \nonumber \\
 &&
 - 2 \left( 1- \frac{1}{\HH r} \right) \ndv \left( \frac{\partial \Delta^{(1)}}{\partial \ln L} \right)_{L = \bar L} \, ,
\\
\Delta^{(3)} \left( \bn , z , >L \right)
&\simeq&\Delta^{(3)} \left( \bn , z , >\bar L \right)
+ 2 \delta d^{(1)} \left( \frac{\partial \Delta^{(2)}}{\partial \ln L} \right)_{L = \bar L}
+2 \delta d^{(2)} \left( \frac{\partial \Delta^{(1)}}{\partial \ln L} \right)_{L = \bar L}
\nonumber \\
&&
- 2 Q \left(  \delta d^{(3)}+ \delta d^{(2)} \Delta^{(1)}  +\delta d^{(1)} \Delta^{(2)}  \right)
\nonumber \\
&\simeq&\Delta^{(3)} \left( \bn , z , >\bar L \right)
- 2 \left( 1- \frac{1}{\HH r} \right) \ndv \left( \frac{\partial \Delta^{(2)}}{\partial \ln L} \right)_{L = \bar L}
\nonumber \\
&&
- 2 \left( 1- \frac{1}{\HH r} \right) \ndvtwo   \left( \frac{\partial \Delta^{(1)}}{\partial \ln L} \right)_{L = \bar L}
\nonumber \\
&&
+ 2 Q \left( 1- \frac{1}{\HH r} \right) \left\{  \ndvthree+ \ndvtwo \left( \delta_g + \frac{1}{\HH}\partial_r \ndv \right) 
\right.
\nonumber \\
&&
\left.
 +\ndv \left( \delta_g^{(2)} + \HH^{-1} \partial_r \ndv^{(2)} 
+ \HH^{-1} \partial_r \left( \ndv  \delta_g \right)  + \HH^{-2}  \partial_r \left( \ndv \partial_r \ndv \right)\right) \right\}  \, ,
\eea
where we have defined 
\be
Q \equiv -  \left( \frac{\partial  \ln \bar n}{\partial \ln L} \right)_{L = \bar L} = \frac{5}{2} s \, .
\ee
By considering that the dark matter density and velocity perturbations are independent from the luminosity, we have
\bea
 \left( \frac{\partial \Delta N^{(1)}}{\partial \ln L} \right)_{L = \bar L} &\simeq& \delta \left( \frac{\partial b_1 }{\partial \ln L} \right)_{L = \bar L} \, , 
 \\
\left( \frac{\partial \Delta N^{(2)}}{\partial \ln L} \right)_{L = \bar L}&\simeq&
 \delta^{(2)} \left( \frac{\partial b_1 }{\partial \ln L} \right)_{L = \bar L} + \frac{1}{2} \delta^2   \left( \frac{\partial b_2 }{\partial \ln L} \right)_{L = \bar L} + \left( K_{ij} \right)^2   \left( \frac{\partial b_{K^2} }{\partial \ln L} \right)_{L = \bar L}
 \nonumber \\
 &&
 + \HH^{-1} \partial_r \left( \ndv \delta \right)   \left( \frac{\partial b_1 }{\partial \ln L} \right)_{L = \bar L} \, .
\eea

\section{Dipole kernels}
\label{app:dipole_kernels}
We define explicitly the function introduced in Eqs.~(\ref{P22dip}-\ref{P13dip}) for the next-to-leading order dipole, where\footnote{To avoid confusion we indicates with $R$ the comoving distance through Eqs.~(\ref{first_eq_appD}-\ref{last_eq_appD})} $r=q/k$ and $x=\hat \bk \cdot \hat \bq$:
\bea
\label{first_eq_appD}
J_{22}^{\Delta b_1} \left( r, x \right) &=&
\frac{9 f^2 \left(r^2 \left(12 x^4-2 x^2-5\right)+2 r x \left(6-11 x^2\right)+8 x^2-3\right)}{140 r^2 \left(r^2-2 r x+1\right)^2}  \, ,
\nonumber \\
&&
+\frac{f}{140 r^2 \left(r^2-2 r x+1\right)^2} 
\left[ 32 r^4 \left(2 x^2+1\right)-32 r^3 x \left(4 x^2+5\right)
\right.
\nonumber
\\
&&
\left.
\qquad
+r^2 \left(6 x^2-1\right) \left(38 x^2+37\right)+6 r x \left(19-64 x^2\right)+135 x^2-48\right]
\nonumber 
\\
&&
+\frac{\left(2 r^2-4 r x+3 x^2-1\right) \left(r \left(2 r x^2+r-3 x\right)+1\right)}{7 r^2 \left(r^2-2 r x+1\right)^2} \, ,
\\
J_{22}^{\Delta b_2} \left( r, x \right) &=&
-\frac{3 f \left(4 r \left(2 r x^2+r-3 x\right)+3\right)}{20 r^2 \left(r^2-2 r x+1\right)} \, ,
\\
J_{22}^{\Delta b_{12}} \left( r, x \right) &=&
-\frac{3 \left(r \left(2 r x^2+r-3 x\right)+1\right)}{4 r^2 \left(r^2-2 r x+1\right)} \, ,
\\
I_{22}^{\Delta b_{1}} \left( r, x \right) &=&\frac{f^3}{35 \HH r^2 {R} \left(r^2-2 r x+1\right)^2} \left(2 r^2 x^4 (\HH (5-6 {\mathcal{R}}) {R}+6)
\right.
\nonumber \\
&&
\qquad\qquad
-\left(r^2+2\right) x^2 (\HH (6 {\mathcal{R}}-5) {R}-6)+3 r^2 (\HH ({\mathcal{R}}-2) {R}-1)
\nonumber \\
&&
\left.
\qquad\qquad 
+5 r x^3 (\HH (6 {\mathcal{R}}-5) {R}-6)+7 \HH r {R} x-3 \HH {\mathcal{R}} {R}-\HH {R}+3\right)
\nonumber \\
&&
+\frac{f^2 \left(2 r x^2+r-3 x\right) \left(r \left(8 r^2-16 r x+42 x^2-13\right)-21 x\right) (\HH (2-3 {\mathcal{R}}) {R}+2)}{210 \HH r^2 {R} \left(r^2-2 r x+1\right)^2}
\nonumber \\
&&
-\frac{f {\mathcal{R}} \left(r \left(10 x^2-3\right)-7 x\right) \left(r \left(16 r^2-32 r x+42 x^2-5\right)-21 x\right)}{294 r^2 \left(r^2-2 r x+1\right)^2}  \, ,
\\
I_{22}^{\Delta b_{2}} \left( r, x \right) &=&
\frac{f^2 \left(2 r x^2+r-3 x\right) ( (3 {\mathcal{R}}-2) {\HH} R -2)}{10 r {\HH} \left(r^2-2 r x+1\right) R}+\frac{f \left(6 r x^2+r-7 x\right) {\mathcal{R}}}{14 r \left(r^2-2 r x+1\right)}  \, ,
\\
I_{22}^{\Delta b_{12}} \left( r, x \right) &=&
\frac{f \left(r \left(2 x^2-1\right)-x\right) {\mathcal{R}}}{2 r \left(r^2-2 r x+1\right)} \, ,
\\
J_{13}^{\Delta b_1} \left( r \right) &=& 
\frac{3 f^2}{70 r^2}+
\frac{f \left(-6 r^7-32 r^5-98 r^3+3 \left(r^2-1\right)^3 \left(r^2+8\right) \log \left(\frac{r+1}{\left| 1-r\right|}\right)+48 r\right)}{560 r^5} \, , \qquad 
 \\
J_{13}^{\Delta b_2} \left( r \right) &=&\frac{9 f}{10 r^2} \, ,
 \\
J_{13}^{\Delta b_{12}} \left( r \right) &=&\frac{3}{2 r^2} \, ,
\\
I_{13}^{\Delta b_1} \left( r \right) &=& 
f^2 \frac{3 {\HH} R \left( \mathcal{R} -2 \right)-2}{280 r^5 {\HH} R}
 \left(3 \left(r^2+2\right) \left(r^2-1\right)^3 \log \left(\frac{\left| r-1\right| }{r+1}\right) \right.
 \nonumber \\
 &&
 \qquad \qquad  \qquad  \qquad  \qquad
 \left.
 +2 r \left(3 r^6-2 r^4+41 r^2-6\right)\right)
\nonumber \\
&&
+\frac{f {\mathcal{R}} \left(3 \left(2 r^2+1\right) \left(r^2-1\right)^3 \log \left(\frac{\left| r-1\right| }{r+1}\right)+2 r \left(6 r^6-13 r^4+30 r^2-3\right)\right)}{168 r^5}
\nonumber \\
&&
+\frac{2 f^3 ({\HH} R (3 {\mathcal{R}}-1)-3)}{15 r^2 {\HH} R} \, ,
\\
K_{13}^{\Delta b_1}  &=&\frac{22}{189} f \mathcal{R} \, ,
\\
K_{13}^{\Delta b_2}  &=&-\frac{10}{7}  f {\mathcal{R} } \, ,
\\
K_{13}^{\Delta b_{3}}  &=&-\frac{1}{2} f {\mathcal{R} } \, .
\label{last_eq_appD}
\eea

\section{Velocity dispersion}
\label{app:veldisp}

We remark that linear theory underestimates the velocity dispersion
\be \label{veldisp}
\sigma_v^2 \equiv \int \frac{d^3q}{\left( 2 \pi \right)^3} \frac{P\left( q \right)}{q^2} \, .
\ee
For instance, the velocity dispersion predicted by linear theory is suppressed with respect to Halofit~\cite{Takahashi:2012em} by a factor $\sim 0.74$ at low redshift, because the non-linear spectrum enhances power on small scales with respect to the linear power spectrum. To account for this effect we introduce an additional EFT-inspired parameter $\sigma_0$ through
\be
\sigma_v^2 \rightarrow \sigma_v^2+ \sigma_0^2 \, .
\ee
In our work we fit for $\sigma_0$ by comparing with the simulations of Ref.~\cite{Breton:2018wzk}.

\begin{figure}[h!]
\begin{center}
\includegraphics[width=0.45\textwidth]{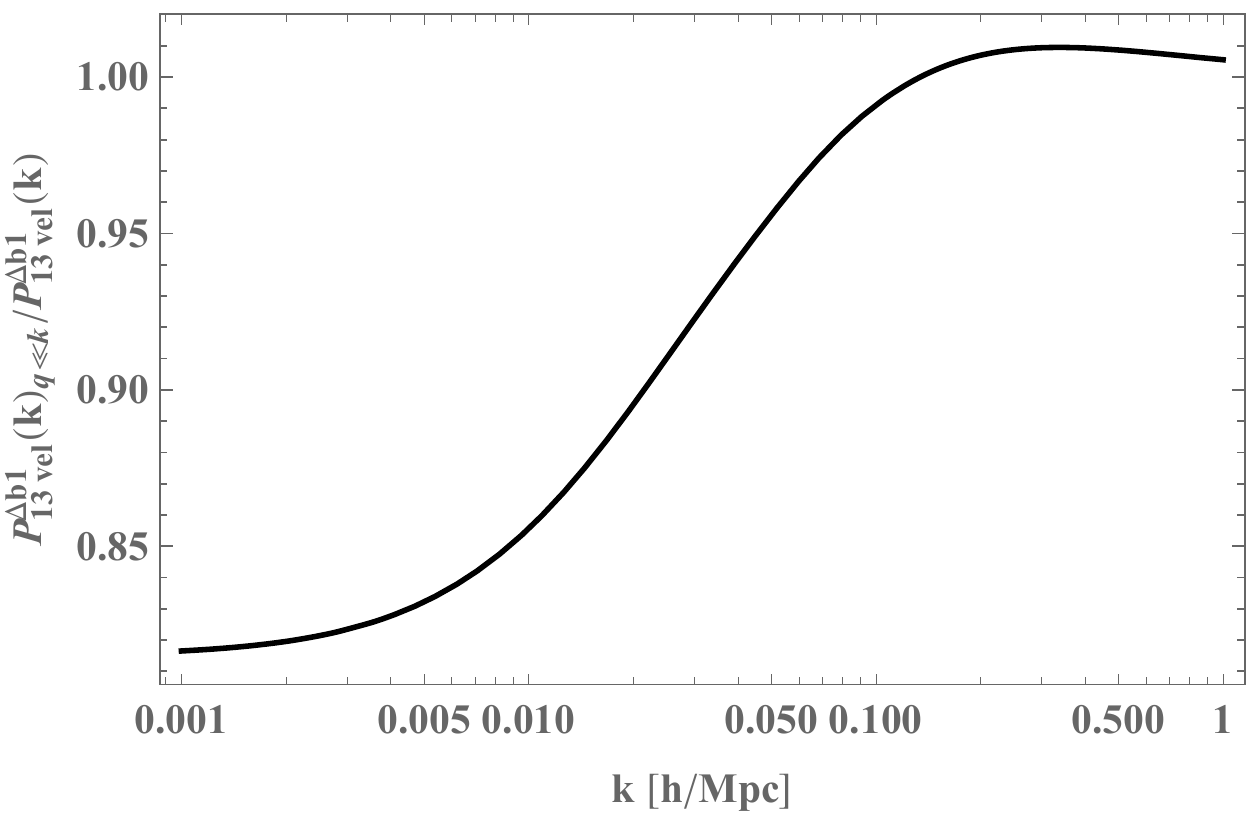}
\caption{We show the ratio between the contribution to the dipole of the kernel $I_{13}^{\Delta b_1} $ compared to its limit solution~\eqref{eq_limsol}.} 
\label{fig_13limit}
\end{center}
\end{figure}

The kernels $J_{13}^{\Delta b_2} $ and $J_{13}^{\Delta b_{12}} $ are directly proportional to $\sigma_v^2$, therefore it is enough to multiply their contributions by the ratio $\left( \left( \sigma^{\rm Lin}_v\right)^2 +\left( \sigma^{\rm Sim}_0\right)^2 \right) /\left( \sigma^{\rm Lin}_v\right)^2 \sim 2 $.
The kernels $J_{13}^{\Delta b_1} $ and $I_{13}^{\Delta b_{1}} $ have a less trivial scale-dependence. From fig.~\ref{fig_correlation} we notice that the contribution of $J_{13}^{\Delta b_1} $ is completely negligible, and so we focus on $I_{13}^{\Delta b_{1}} $ only.  As we show in fig.~\ref{fig_13limit}, its behaviour in the mildly nonlinear regime is fully captured by its limit solution
\be
\label{eq_limsol}
I_{13}^{\Delta b_1} \rightarrow
\frac{f^3 ((6\mathcal{R}-2) \HH R-6)}{15 r^2 \HH R}+\frac{ f^2 ((3 \mathcal{R}-2) \HH R-2)}{5 r^2 \HH R}+\frac{ \mathcal{R} f}{3 r^2}
\quad \text{for} \quad q\ll k \, .
\ee 
Hence, every kernel which contributes to $P_{13}$ can be simply considered to be proportional to the velocity dispersion~\eqref{veldisp}.
In figs.~\ref{fig_correlation_tot}~and~\ref{fig_correlation_tot_NoEuler} we extrapolate the value of $\sigma^{\rm Sim}_0$ to higher redshifts than the simulations of Ref.~\cite{Breton:2018wzk} by considering the redshift evolution of the velocity dispersion. Therefore, this stands as a first approximation due to the lack of relativistic simulations available at those redshifts.

\newpage
\bibliographystyle{JHEP}
\bibliography{biblio_dipole}

\end{document}